\documentclass[12pt]{article}
\usepackage{amssymb}
\usepackage[dvips]{graphics}
\usepackage{epsfig}
\newcommand{\be}{\begin{equation}}
\newcommand{\ee}{\end{equation}}
\newcommand{\ii}{\textrm{i}}

\begin{document}
\title{\LARGE{\textbf{Critical exponents for the $\textrm{FPL}^2$ model}}}
\author{\small{David Dei Cont and Bernard Nienhuis}}
\date{\normalsize \emph{Instituut voor Theoretische Fysica, Universiteit van Amsterdam,\\
Valckenierstraat 65, 1018 XE Amsterdam, The Netherlands} \\
deicont@science.uva.nl, nienhuis@science.uva.nl
}
\maketitle
\thispagestyle{empty}
\begin{abstract}
Starting from the Bethe ansatz solution we derive
a set of coupled non-linear integral equations for the
fully packed double loop model ($\textrm{FPL}^2$)
on the square lattice.
As an application we find exact expressions for the
central charge and for the scaling dimension corresponding
to the simplest charge excitation.
We study numerically the low-lying excitations
corresponding to more general perturbations of the ground state
and discover that the corresponding scaling dimensions
are well described by the Cartan matrix of $s \ell_{4}$.
\\ 
\end{abstract}
\clearpage
\section{Introduction}
The fully packed double loop model ($\textrm{FPL}^{2}$)
is a two dimensional statistical model constructed by 
decorating the square lattice with two species of loops
(black and grey), under the constraint that every lattice
edge is covered by only one loop
and every vertex is visited by both types of loops.

It exhibits a rich phase diagram and provides a
representation for previously studied models:
Ice model \cite{fsL1}, four-colouring model \cite{fsREAD,fsKH1,fsKH2,BERNARD},
dimer loop model \cite{fsRAGHAVAN}, double Hamiltonian walk 
\cite{fsKONDEV4}
and compact polymers \cite{fsBatchelor2}.
A generalization of the $\textrm{FPL}^2$ model, which
offers a unifying picture of the compact, dense and dilute
phase of polymers, was proposed in \cite{fsJ3} by relaxing
the full packing constraint.
Another possible generalization corresponds 
to the Flory model of polymer melting \cite{fsJ2,fsJ2bis}.

Jacobsen and Kondev mapped the $\textrm{FPL}^{2}$
model onto a height model and postulate that the
long wavelength behaviour is correctly described
by a Liouville field theory.
Making use of the Coulomb gas method they succeed
in computing the central charge and the critical exponents \cite{fsJK1,fsKONDEV3}.
The study of the relation between the Coulomb gas 
representation and the underlying 
conformal field theory has been initiated in \cite{DOTSENKO}.
 
Recently we mapped the $\textrm{FPL}^{2}$ model
onto a 24-vertex model and diagonalized
the corresponding transfer matrix by means
of coordinate Bethe ansatz (BA) \cite{D,BERNARD}.
The solution holds only when the two loop
fugacities are the same.
In particular we calculated analytically
the configurational entropy for the double Hamiltonian
walk and the four-coloring model.
We also confirmed a conjecture on the average length of loops \cite{fsJ1}.
In the particular limit in which the fugacities of the loops
are both equal to two, the BA equations become rational \cite{BERNARD}
and look very much like those derived in \cite{M}
for mixed $\textrm{SU}(N)$ vertex model for the case $N=4$.
 
When we made this paper ready for publication,
Jacobsen and Zinn-Justin \cite{fsJZ} succeed in finding        
a Yang-Baxter structure, for the $\textrm{FPL}^{2}$ model, in
terms of the affine quantum group $U_{q}(\widehat{s\ell(4)})$.
They reobtain the BA equations following the algebraic approach
and write exact expressions for the central charge and for 
the scaling dimensions of the low-lying excitations.

These quantities can be calculated from the leading 
finite-size corrections to the ground \cite{fsBCN,fsIA}
and the excited \cite{fsCardy1,fsCardy2} energies.
Analytical calculations, for the low-lying excitations
of models solvable by the Bethe ansatz equations,
were done in \cite{fsBogoliubov,fsIzergin}.
An analytical method, based on the construction
of a non-linear integral equation (NLIE) for the
counting function, which allows the calculation
of the central charge, has been developed in 
\cite{fsPK,fsDDV3}.
This method has been exploited in
\cite{fsPK,fsDDV1,fsDDV2,fsDT1,fsDT2,fsFioravanti1}
in order to include excitations.
The case of multicomponent BA equations
has been investigated in \cite{fsPZJ1}.

Here, following \cite{fsPZJ1}, 
we write a non-linear integral equation for the
$\textrm{FPL}^2$ model in the presence of the seam.
As an application we compute analytically the
value of the central charge and of the scaling dimension
corresponding to one grey and one black string
propagating between two points of the lattice.
Our results are in agreement with 
the results found in \cite{fsJK1} via the field theory.
We also studied numerically the low-lying excitations 
\cite{fsKalugin}
generated by varying the number of particles 
(charge excitations), allowing transitions 
from one side to the other of the Fermi seas 
(umklapp processes) and perturbing the surfaces
of the Fermi seas (particle-hole excitations) 
\cite{fsBogoliubov}.
It turns out that the scaling dimensions are well
described by a compact formula \cite{fsIzergin} 
through the Cartan matrix of $s \ell_{4}$.
Our results coincide with those found in \cite{fsJZ}.

The present work is organized as follows.
In section \ref{sectionMODEL} we introduce the 
$\textrm{FPL}^2$ model, summarize the main
results of \cite{D} and define the counting functions.
For completeness in section \ref{sectionSUMINTEGRAL}
we show in detail how the finite sums entering 
in the definition of the counting functions
can be expressed in terms of integrals involving
the counting functions.
This procedure is well known in the literature.
See for instance \cite{fsDDV2,fsPZJ1}.
In section \ref{sectionNLIE} we derive a set of coupled
non-linear integral equations for the counting functions,
in the presence of the seam.
An integral expression for the finite-size free energy
is derived in section \ref{sectionENERGY}.
The central charge is calculated in section \ref{sectionCHARGE}
and the scaling dimension associated 
with one grey and one black string derived in section
\ref{sectionEXCITATION}.
In section \ref{sectionNUMERICS} we study numerically
the low-lying excitations.
%
\section{The model \label{sectionMODEL}}
\begin{figure}[!b]
\centerline{\epsfxsize=15cm \epsfbox{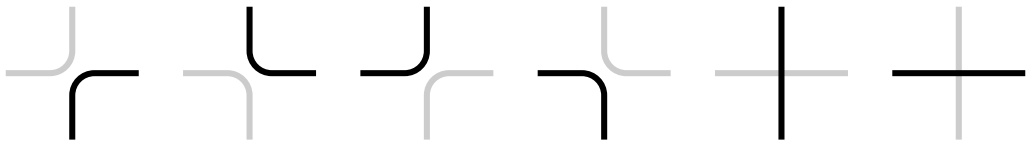}}
\caption{The six vertex configurations of the
$\textrm{FPL}^2$ model that are allowed by the
fully-packing constraint.
Each vertex is visited by both types of loops.
The two rightmost vertices permit the two species
to cross.
If we assign an orientation to the loops we obtain
24 distinct vertices.
}
\label{fig1}
\end{figure}
The fully packed double loop model ($\textrm{FPL}^2$) 
is a two dimensional statistical model constructed 
decorating the square lattice with two species of loops.
Each bond of the square lattice is covered by only one
loop and every vertex is visited by both types of loops.
Loops of the same type are not allowed to cross.
Representing the two species by black 
and grey segments respectively,
the fully-packing constraint forces each vertex of the
square lattice to have one of the six appearances
depicted in figure~\ref{fig1}.
We place the model on a cylinder of circumference
$L$ and height $M$ with periodic boundary conditions
also in the vertical direction.
The partition function is defined by:
\be
\label{fspartitionsum}
Z(L,M)
=
\sum 
n_{b}^{N_{b}} n_{g}^{N_{g}}
m_{b}^{M_{b}} m_{g}^{M_{g}}
\ee
where the sum runs over all the allowed configurations.
The exponents $N_{b}$ and $N_{g}$ are the respective 
numbers of black and grey contractable loops.
Since the model is placed on a cylinder loops
are allowed to wind.
We denote by $M_{b}$ and $M_{g}$ the number
of black and grey winding loops.
We assign the fugacities $n_{b,g}$ for contractable
loops and $m_{b,g}$ for uncontractable ones.

In \cite{D} we map the loop model into a 24-vertex model
and construct the transfer matrix $\mathbf{T}$.
The partition function is the sum of the eigenvalues 
$\Lambda_{i}$
of the transfer matrix, raised to the power $M$:
\be
Z(L,M)=\textrm{Tr} \, \mathbf{T}^M(L) = \sum \Lambda_{i}^{M}(L).
\ee
In the limit in which the height of the cylinder extend to
infinite $M \rightarrow \infty$ the finite-size free
energy density $f(L)$ can be expressed in terms of the largest
eigenvalue $\Lambda_{0}(L)$ of the transfer matrix:
\be
\label{fsFL}
f(L)=-\frac{1}{L} \log \Lambda_{0}(L).
\ee
At criticality, the dominant finite-size correction
of the free energy density,
can be related to the central charge $c$ and the 
bulk free energy $f(\infty)$ via the formula 
\cite{fsBCN,fsIA}:
\be
\label{caricacentrale}
f(L)=f(\infty)-\frac{c \, \pi}{6 \, L^2}+o(1/L^2).
\ee
The subdominant eigenvalues $\Lambda_{i}$ 
allow the computation of the scaling dimensions
$\Delta_{i}$ thanks to the 
formula \cite{fsCardy1,fsCardy2}:
\be
\label{fsDELTA}
\log 
\frac{\Lambda_{0}(L)}{\Lambda_{i}(L)}
=
\frac{2 \, \pi}{L} \, \Delta_{i} + o(1/L).
\ee

In \cite{D} we show that the model is solvable
in the particular case in which the fugacities
are the same ($n_{b}=n_{g}$ and $m_{b}=m_{g}$).
The transfer matrix is diagonalized using
coordinate nested Bethe ansatz and an analytical
expression for the free energy, in the 
thermodynamic limit, is derived.
In particular we find the free energy of the
four-coloring model \cite{fsKH1,fsKH2,BERNARD}
and the double Hamiltonian walk \cite{fsKONDEV4}
and recover the known entropy 
of the Ice model \cite {fsL1}.
Let us summarize the main result of $\cite{D}$:
The fugacities entering in the partition sum
(\ref{fspartitionsum}) are expressed in terms
of two phases $\theta$ and $\alpha$,
through the relations:
\be
n_{b}=n_{g}=2 \, \cos \pi \, \theta
\qquad
m_{b}=m_{g}=2 \, \cos \pi \, \alpha
\qquad
0 \leq \theta, \alpha \leq 1
\ee
and a $\theta$-dependent function is introduced:
\be
\label{fsrational}
S_{c}(x,\theta)
\equiv
\frac
{
\sin\frac{\pi \, \theta}{2}(c+x \, \textrm{i})
}
{
\sin\frac{\pi \, \theta}{2}(c-x \, \textrm{i})
}
.
\ee
With these definitions the Bethe 
ansatz equations (BAE) assume the form:
\be
\label{fsAlgebraicBA1}
S_{1}(u_{j},\theta)^{L/2}
=
-
\prod_{i=1}^{n_{w}} S_{1}(w_{i}-u_{j},\theta)
\prod_{k=1}^{n_{u}} -S_{2}(u_{j}-u_{k},\theta)
\ee
\be
S_{1}(v_{j},\theta)^{L/2}
=
-
\prod_{i=1}^{n_{w}} S_{1}(w_{i}-v_{j},\theta)
\prod_{k=1}^{n_{v}} -S_{2}(v_{j}-v_{k},\theta)
\ee
\be
\label{fsAlgebraicBA2}
\textrm{exp}(-2 \, \textrm{i} \, \pi \, \alpha)
\prod_{j=1}^{n_{u}} S_{1}(w_{l}-u_{j},\theta)
\prod_{k=1}^{n_{v}} S_{1}(w_{l}-v_{k},\theta)
\prod_{m=1}^{n_{w}} -S_{2}(w_{m}-w_{l},\theta)
=
-1
.
\ee
Every solution of the previous equations,
with non coinciding roots, is in correspondence 
with an eigenvalue of the transfer
matrix via the expression:
\begin{eqnarray}
\label{fsAUTOVALORE}
\Lambda
& = &
(-1)^{n_{w}}
\left[
\textrm{exp}(\textrm{i} \, \pi \, \alpha)
\prod_{j=1}^{n_{u}} S_{1}(u_{j},\theta)^{1/2}
\prod_{k=1}^{n_{v}} S_{1}(v_{k},\theta)^{1/2}
\prod_{m=1}^{n_{w}} 
\frac{\sin \frac{\pi \theta}{2}(w_{m} \, \textrm{i}-2)}{\sin \frac{\pi \theta}{2} w_{m} \, \textrm{i}}
\right.
\\ \nonumber
& + &
\left.
\textrm{exp}(-\textrm{i} \, \pi \, \alpha)
\prod_{j=1}^{n_{u}} S_{1}(u_{j},\theta)^{-1/2}
\prod_{k=1}^{n_{v}} S_{1}(v_{k},\theta)^{-1/2}
\prod_{m=1}^{n_{w}} 
\frac{\sin \frac{\pi \theta}{2}(w_{m} \, \textrm{i}+2)}{\sin \frac{\pi \theta}{2} w_{m} \, \textrm{i}}
\right].
\end{eqnarray}

In \cite{fsJK1} the central charge and a set of
scaling dimensions were computed analytically
but not rigorously via the field theory
making use of Coulomb gas techniques.
In this paper we are going to compute the conformal
quantities starting from the Bethe ansatz solution.
These results are included in the independent
and recent paper \cite{fsJZ}.

The calculation is based on the formula 
(\ref{caricacentrale}) and (\ref{fsDELTA}) that
relate the central charge and the scaling dimensions
to the leading finite-size correction of the ground
and excited energies.
An analytical method, based on the perturbation of the
Fermi sea, was developed for the first time in
\cite{fsBogoliubov,fsIzergin}.
Later another method, based on the construction of
a non-linear integral equation, was developed.
Following this method the central charge, of models
solvable by Bethe ansatz equations, was calculated
in \cite{fsPK,fsDDV3} and the scaling dimensions
in \cite{fsPK,fsDT1,fsDT2,fsFioravanti1}.
The method has been refined to include string excitations
\cite{fsDDV2} and multicomponents BA equations \cite{fsPZJ1}.
Here we are going to follow this last approach.
The central objects which enter in this machinery
are the counting functions $Z_{L,s}$ defined by:
\begin{eqnarray}
\nonumber
Z_{L,u}(u,\theta) 
& \equiv & 
-\frac{L}{2} \phi_{1}(u,\theta)
+
\sum_{m=1}^{n_{w}} \phi_{1}(w_{m}-u,\theta)-\sum_{k=1}^{n_{u}} \phi_{2}(u_{k}-u,\theta)
\\  
\label{fscountfunc}
Z_{L,w}(w,\theta) 
& \equiv & 
\sum_{j=1}^{n_{u}} \phi_{1}(u_{j}-w,\theta)
+
\sum_{k=1}^{n_{v}} \phi_{1}(v_{k}-w,\theta)
-
\sum_{l=1}^{n_{w}} \phi_{2}(w_{l}-w,\theta)
-
2 \, \pi \, \alpha
\\ \nonumber
Z_{L,v}(v,\theta) 
& \equiv &
-\frac{L}{2} \phi_{1}(v,\theta)
+
\sum_{m=1}^{n_{w}} \phi_{1}(w_{m}-v,\theta)-\sum_{k=1}^{n_{v}} \phi_{2}(v_{k}-v,\theta)
\end{eqnarray}
where the following function has been defined:
\be
\label{fsphic}
\phi_{c}(x,\theta) 
\equiv
\ii \, \log S_{c}(x,\theta).
\ee
Notice that the counting functions 
are constructed by taking the logarithm
of the equations 
(\ref{fsAlgebraicBA1}-\ref{fsAlgebraicBA2}).
The key property of $Z_{L,s}$ is that it counts
the BA roots, in the sense that for each root
$\lambda_{s,k}$ of the BAE 
(\ref{fsAlgebraicBA1}-\ref{fsAlgebraicBA2}) we have:
\be
\label{fsCONTA}
Z_{L,s}(\lambda_{s,k})
=
2 \, \pi \, \textrm{I}_{s,k}
\qquad
k=1,\ldots,n_{s}
\qquad
s=u,w,v
\ee
where $\textrm{I}_{s,k}$
is an integer or half-integer number, 
depending on the number of roots of each type.
We will call holes, objects $h_{s,k}$ that are
counted by $Z_{L,s}$:
\be
Z_{L,s}(h_{s,k})=2 \, \pi \, \textrm{I}_{H,s,k}
\qquad
k=1,\ldots,n_{s}
\qquad
s=u,w,v
\ee
but are not solution of the system of BA equations 
(\ref{fsAlgebraicBA1}-\ref{fsAlgebraicBA2}).
Where $\textrm{I}_{H,s,k}$ are integers 
or half-integer numbers.
The first step towards the exact calculation
of the finite-size corrections consists in the
replacement of the \emph{finite} sums 
in the rhs of (\ref{fscountfunc}) by integrals.
In the next section we are going to show
how this is possible.
\section{From a finite sum to an integral \label{sectionSUMINTEGRAL}}
We want to find an integral expression for the finite sums
entering in the counting functions (\ref{fscountfunc}) 
for the particular case in which the BA roots $\lambda_{s,k}$
are positioned on the real line and no real holes are present.
These sums have the general form:
\be
\label{lasomma}
\sum_{k=1}^{n_{s}}
\phi_{c}(\lambda_{s,k}-\lambda,\theta).
\ee
Start by noting that with the help of the counting function
it is possible to construct a function which is analytic in all
the complex plane and has zeros of order one at the BA roots.
In fact, the following expression satisfies the requirements:
\be
\label{fsDETECT}
1+(-1)^{\delta} \, \textrm{exp}[\textrm{i} \, Z_{L,s}(z)]
\ee
where we choose $\delta=1$ for the case in which the number of roots
is such that $Z_{L,s}(\lambda_{s,k})/\pi$ is even and $\delta=0$ when it is odd.
Applying Cauchy formula we can prove the identity:
\be
\label{fsidentity} 
2 \, \pi \, \ii \, \sum_{k=1}^{n_{s}} \phi_{c}(\lambda_{s,k}-\lambda,\theta)
=
\int_{\Gamma} d\mu \, 
\phi_{c}(\mu-\lambda,\theta) \, 
\frac{\ii  Z^{\prime}_{L,s}(\mu) \, (-1)^{\delta} \, \textrm{exp}[\ii  Z_{L,s}(\mu)] }
{1+(-1)^{\delta} \, \textrm{exp}[\ii  Z_{L,s}(\mu)]}
\ee
where $\Gamma$ is a close contour which encircles 
counterclockwise the BA roots  $\lambda_{s,k}$.
A fundamental assumption for the validity of the previous equality
is the analyticity of $\phi_{c}(z,\theta)$ inside and on the contour $\Gamma$.
In the following we will use the shorthand notation:
\be
\label{Rdelta}
R_{s,\delta}(z)
\equiv
\frac{\ii \, Z^{\prime}_{L,s}(z) \, (-1)^{\delta} \, \textrm{exp}[\ii Z_{L,s}(z)] }
{1+(-1)^{\delta} \, \textrm{exp}[\ii Z_{L,s}(z)]}.
\ee
We assumed that the roots of the BA equations were located on the real axis.
Therefore we choose for $\Gamma$ a box centered
at the origin, extending from  $-\infty$ to $+\infty$
in the horizontal direction and of height $2 \, \eta$,
and integrate counterclockwise.
In the limit $\eta \rightarrow 0^{+}$ only the integrals
running on the side parallel to the real axis survive
and the rhs of (\ref{fsidentity}) becomes:
\begin{eqnarray}
\label{RDminRDplus}
\int_{\Gamma} d\mu \, \phi_{c}(\mu-\lambda,\theta) \, R_{s,\delta}(\mu)
& = & 
\lim_{\eta \rightarrow 0^{+}}
\Big[
\int_{-\infty}^{+\infty} dx \, \phi_{c}(x-\ii\eta-\lambda,\theta) \, R_{s,\delta}(x-\ii\eta)
\\ \nonumber
& + &
\int_{+\infty}^{-\infty} dx \, \phi_{c}(x+\ii\eta-\lambda,\theta) \, R_{s,\delta}(x+\ii\eta)
\Big].
\end{eqnarray}
The next step consists in the manipulation of the expressions
$R_{s,\delta}(x + \ii \, \eta)$ and $R_{s,\delta}(x - \ii \, \eta)$.
At a first glance one would say that:
\be
\label{RLOG}
R_{s,\delta}(x \pm \ii \eta)
=
\frac{d}{d x} \log \Big[ 1+(-1)^{\delta} \, \textrm{exp}[\ii Z_{L,s}(x \pm \ii \eta)] \Big].
\ee
But this is not always true because
the logarithm has a cut on the negative real line.
We analyze this point in more details focusing
on the term $R_{s,\delta}(x + \ii \, \eta)$.
Let us study the behaviour of the argument of the
logarithm in a neighborhood of a 
solution $x$ of the BA equations.
To construct the neighborhood take a small real number
$\epsilon$ and remind that $\eta > 0$.
The following expansion hold:
\be
\label{Rexpansion}
1+(-1)^{\delta} \, \textrm{exp}[\ii \, Z_{L,s}(x + \epsilon+\ii \, \eta)]
=
(\eta-\ii \, \epsilon) \, Z^{\prime}_{L,s}(x) + \mathcal{O}((\epsilon+\ii \, \eta)^2).
\ee
From this expansion we see that if $Z^{\prime}_{s,L}(x)>0$ 
the argument of the logarithm has a positive real part.
Thus varying $\epsilon$ it will never cross 
the cut and the equality (\ref{RLOG}) is satisfied.
On the other hand if $y_{s,k}$ is a solution of the
BAE for which $Z^{\prime}_{L,s}(y_{s,k})<0$ 
(we call $y_{s,k}$ a special object) the logarithm 
will exhibit a discontinuity at $\epsilon=0$.
However the expression $R_{s,\delta}(x+\ii \eta)$
is continuous when $x$ runs from $-\infty$ to $+\infty$
and so it must be the derivative of a continuous function.
Thus in the presence of special objects $y_{s,k}$
relation (\ref{RLOG}) must be replaced 
with the following equality:
\be
\label{fsspecialobject1}
R_{s,\delta}(x+\ii \eta)
= 
\frac{d}{d x} 
\log \Big[ 1+(-1)^{\delta} \, \textrm{exp}[\ii Z_{L,s}(x+\ii\eta)] \Big]
+
2 \pi \ii \sum_{k=1}^{n_{y_{s}}}\delta(x-y_{s,k}).
\ee
It can be proved that a similar relation holds for 
$R_{s,\delta}(x-\ii \eta)$:
\be
\label{fsspecialobject2}
R_{s,\delta}(x-\ii \eta)
=
\ii \, Z_{L,s}^{\prime}(x-\ii \eta)
+
\frac{d}{d x}
\log
\Big[
1
+
(-1)^{\delta} \, \textrm{exp}[-\ii Z_{L,s}(x-\ii\eta)]
\Big]
-
2 \pi \ii \sum_{k=1}^{n_{y_s}}\delta(x-y_{s,k}).
\ee
Comparing this equality with (\ref{fsspecialobject1})
we see that the special objects 
contribute with the opposite sign and 
that the extra term $Z^{\prime}(x-\ii \eta)$ arises.
Substituting (\ref{fsspecialobject1}) 
and (\ref{fsspecialobject2}) 
into (\ref{RDminRDplus}) yields:
\begin{eqnarray}
\lefteqn{
\int_{\Gamma} d\mu \, \phi_c(\mu-\lambda,\theta) \, R_{s,\delta}(\mu)
=
\ii
\int_{-\infty}^{+\infty} \, dx \, \phi_{c}(x-\ii\eta-\lambda,\theta) Z_{L,s}^{\prime}(x-\ii\eta)
\nonumber
}
\\
& + &
\int_{-\infty}^{+\infty} \, dx \, \phi_{c}(x-\ii\eta-\lambda,\theta)
\frac{d}{d x}
\log
\Big[
1
+
(-1)^{\delta} \, \textrm{exp}[-\ii Z_{L,s}(x-\ii\eta)]
\Big]
\\ \nonumber
& - &
\int_{-\infty}^{+\infty} \, dx \, \phi_{c}(x+\ii\eta-\lambda,\theta)
\frac{d}{d x}
\log
\Big[
1
+
(-1)^{\delta} \, \textrm{exp}[\ii Z_{L,s}(x+\ii\eta)]
\Big]
\\ \nonumber
& - &
2 \, \pi \, \ii
\sum_{k=1}^{n_{y_s}} \phi_{c}(y_{s,k}-\ii \eta-\lambda,\theta)
-
2 \, \pi \, \ii
\sum_{k=1}^{n_{y_s}} \phi_{c}(y_{s,k}+\ii \eta-\lambda,\theta).
\end{eqnarray}
From the definitions (\ref{fsrational},\ref{fsphic},\ref{fscountfunc})
we see that the functions $Z_{L,s}$ and $\phi_{c}$ enjoy the following symmetry
(since $\alpha$ and $\theta$ are real numbers):
\be
\phi_{c}(z,\theta)=\overline{\phi_{c}(\bar{z},\theta)}=-\phi_{c}(-z,\theta)
\qquad
Z_{L,s}(z)=\overline{Z_{L,s}(\bar{z})}
\ee
so that we can write the compact formula:
\begin{eqnarray}
\lefteqn{
\int_{\Gamma} d\mu \, \phi_{c}(\mu-\lambda,\theta) \, R_{s,\delta}(\mu)   
=
\ii
\int_{-\infty}^{+\infty} dx \, \phi_{c}(x-\ii\eta-\lambda,\theta) Z_{L,s}^{\prime}(x-\ii\eta)
}
\\ \nonumber
& - &
2 \, \ii \, \textrm{Im}
\int_{-\infty}^{+\infty} dx \, \phi_{c}(x+\ii\eta-\lambda,\theta)
\frac{d}{d x}
\log
\Big[
1
+
(-1)^{\delta} \, \textrm{exp}[\ii Z_{L,s}(x+\ii\eta)]
\Big]
\end{eqnarray}
where we dropped the terms containing 
the special objects because for the 
finite-size corrections that we will 
investigate they do not appear.
Integrating by parts:
\begin{eqnarray}
\lefteqn{
\int_{\Gamma} d\mu \, \phi_{c}(\mu-\lambda,\theta) \, R_{s,\delta}(\mu)
 = 
\left. \ii \phi_{c}(x-\ii\eta-\lambda,\theta) Z_{L,s}(x-\ii\eta) \right|_{x=-\infty}^{x=+\infty}
\nonumber
}
\\
& - &
\ii
\int_{-\infty}^{+\infty} dx \, \phi_{c}^{\prime}(x-\ii\eta-\lambda,\theta) Z_{L,s}(x-\ii\eta)
\\ \nonumber
& - & 
\left. 2 \, \ii \, \textrm{Im} \, \phi_{c}(x+\ii\eta-\lambda,\theta) 
\log 
\Big[
1+(-1)^{\delta} \, \textrm{exp}[\ii Z_{L,s}(x+\ii\eta)] 
\Big] 
\right|_{x=-\infty}^{x=+\infty}
\\ \nonumber
& + &
2 \, \ii \, \textrm{Im}
\int_{-\infty}^{+\infty} dx \, \phi_{c}^{\prime}(x+\ii\eta-\lambda,\theta) 
\log 
\Big[ 
1+(-1)^{\delta} \, \textrm{exp}[\ii Z_{L,s}(x+\ii\eta)] 
\Big].
\\ \nonumber
\end{eqnarray}
The line of integration of the first integral can be
deformed as long as we stay in the region of analyticity
of the function $\phi_{c}$.
At the end we get the following integral representation
for the finite sum (\ref{lasomma}):
\begin{eqnarray}	
\label{fsKEYRESULT}
\lefteqn{
\sum_{k=1}^{n_{s}}
\phi_{c}(\lambda_{s,k}-\lambda,\theta)
 = 
 - \int_{-\infty}^{+\infty} \frac{dx}{2 \pi} \, \phi_{c}^{\prime}(x-\lambda,\theta) Z_{L,s}(x)
}
\\ \nonumber
& + &
\lim_{\eta \rightarrow 0^{+}}
2  \, \textrm{Im}
\int_{-\infty}^{+\infty} \frac{dx}{2 \pi} \, 
\phi_{c}^{\prime}(x-\lambda,\theta) 
\log 
\Big[ 1+(-1)^{\delta} \, \textrm{exp}[\ii Z_{L,s}(x+\ii\eta)] \Big]
\\ \nonumber
& + &
\textrm{border terms}.
\end{eqnarray}
In the following, for notational convenience,
we will omit the limit symbol and claim
validity of the equations only in the limit 
$\eta \rightarrow 0^{+}$.

Notice that in deriving relation (\ref{fsKEYRESULT})
we made the fundamental assumption that the BA roots 
$\lambda_{s,k}$ were positioned on the real line 
and that real holes were not present.
If real holes $h_{s,k}$ are present the equality (\ref{fsKEYRESULT})
is not valid anymore.
Infact the function (\ref{fsDETECT}) will vanish at the real holes $h_{s,k}$
so that the rhs of (\ref{fsKEYRESULT}) will contain, beside
the contribution coming from the BA roots 
$\sum \phi_{c}(\lambda_{s,k}-\lambda,\theta)$,
also a contribution coming from the BA holes
$\sum \phi_{c}(h_{s,k}-\lambda,\theta)$.
%
\section{Non-linear integral equation \label{sectionNLIE}}
In this section we will derive a set of coupled
non-linear integral equations (NLIE) for the
counting functions.
Making use of the key result (\ref{fsKEYRESULT})
we can reexpress the finite sums that enter in
the rhs of (\ref{fscountfunc}) in terms of 
integrals that involve the counting functions:
\be
\label{fsNLIE1}
\left( \begin{array}{c}
Z_{L,u} \\
Z_{L,w} \\
Z_{L,v}
\end{array}
\right)
=
-
\frac{L}{2} \Phi_{1}
\left( \begin{array}{c}
1 \\
0 \\
1
\end{array}
\right)
+
\mathbf{\Phi}
*
\left( \begin{array}{c}
Z_{L,u} \\
Z_{L,w} \\
Z_{L,v}
\end{array}
\right)
-
2 \, \textrm{Im}\,
\mathbf{\Phi}
*
\left( \begin{array}{c}
Q_{L,u} \\
Q_{L,w} \\
Q_{L,v}
\end{array}
\right)
-
\left( \begin{array}{c}
0                  \\
2 \, \pi \, \alpha \\
0
\end{array}
\right).
\ee
And we have checked that 
the border terms vanish.
Let us clarify the notation:
The symbol $*$ means that in performing the matrix
product the ordinary product has to be replaced with
the convolution product.
We have defined the derivative of (\ref{fsphic}) as:
\be
\Phi_{c}(x,\theta)
\equiv
\frac{1}{2 \, \pi} \, \frac{d}{dx} \phi_{c}(x,\theta)
\ee 
and introduced the shorthand notation for the non-linear part:
\be
Q_{L,s}(x) \equiv \log \Big[1+(-1)^{\delta} \, \textrm{exp}[\ii Z_{L,s}(x+\ii \, \eta)] \Big].
\ee
The matrix $\mathbf{\Phi}$ is defined as:
\be
\mathbf{\Phi}
\equiv
\left( \begin{array}{ccc}
 \Phi_{2}   &  -\Phi_{1} &   0          \\
-\Phi_{1}   &   \Phi_{2} &  -\Phi_{1}   \\
  0         &  -\Phi_{1} &   \Phi_{2}   \\
\end{array}
\right).
\ee
In order to solve equation (\ref{fsNLIE1}) for $Z_{L,s}$
in terms of the non-linear part $Q_{L,s}$ it is convenient to
work with the Fourier transform:
\begin{eqnarray}
\label{fsfourierNLIE1}
\left( \begin{array}{c}
\tilde{Z}_{L,u} \\
\tilde{Z}_{L,w} \\
\tilde{Z}_{L,v}
\end{array}
\right)
& = &
-
\frac{L}{2} \tilde{\Phi}_{1}
\,
(\mathbf{1}-\mathbf{\tilde{\Phi}})^{-1}
\,
\left( \begin{array}{c}
1 \\
0 \\
1
\end{array}
\right)
-
2 \, \textrm{Im} \,
(\mathbf{1}-\mathbf{\tilde{\Phi}})^{-1}
\, 
\mathbf{\tilde{\Phi}}
\,
\left( \begin{array}{c}
\tilde{Q}_{L,u} \\
\tilde{Q}_{L,w} \\
\tilde{Q}_{L,v}
\end{array}
\right)
\\ \nonumber
&  &
-
\delta(p)
\,
(\mathbf{1}-\mathbf{\tilde{\Phi}})^{-1}(0)
\left( \begin{array}{c}	
0       \\
2 \, \pi \, \alpha  \\
0
\end{array}
\right)
\end{eqnarray}
where the Fourier transform and 
its inverse are defined by:
\be
F[\phi](p) 
\equiv 
\tilde{\phi}(p) 
\equiv
\int_{-\infty}^{+\infty} e^{-ipx} \phi(x) dx
\qquad
F^{-1}[\tilde{\phi}](x)
\equiv
\frac{1}{2\pi} \int_{-\infty}^{+\infty} e^{ipx} \tilde{\phi}(p) dp.
\ee
In particular:
\be
\int_{-\infty}^{+\infty} e^{-ipx} \Phi_{c}(x,\theta) dx
=
\frac{\sinh(p(c-1/\theta))}{\sinh(p/\theta)}.
\ee
We will need the following limit:
\be
\label{fsmatrixG}
\mathbf{\tilde{G}}(p,\theta)
\equiv
-
( \mathbf{1}-\mathbf{\tilde{\Phi}} )^{-1} 
\mathbf{\tilde{\Phi}}
,
\qquad
\lim_{p \rightarrow 0}
\mathbf{\tilde{G}}(p,\theta)
=
\frac{1}{4 (1 - \theta)}
\left( \begin{array}{ccc}
1-4 \, \theta &  -2        &    -1         \\
-2        &  -4 \, \theta  &    -2         \\
-1        &  -2        &    1-4 \, \theta  \\
\end{array}
\right).
\ee
%
\begin{figure}[!t]
\centerline{\epsfxsize=10cm \epsfbox{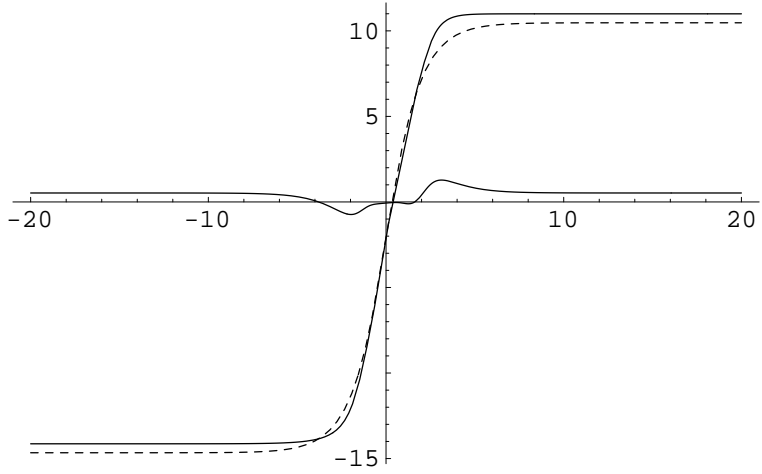}}
\caption{Plot of the terms entering in the second
equation of the non-linear system (\ref{fsNLIE}).
Similar plots hold for the first and the third
equation.
The parameters of the system are 
$L=8$, $n_{u}=n_{w}=n_{v}=L/2$ and
$\theta=\alpha=1/4$.
The curve close to the real axis represents the non-linear
correction $2 \, \textrm{Im} G_{2,s}*Q_{L,s}$ 
which vanishes in the limit $L \rightarrow \infty$.
The dashed line is the bulk term 
$2  \pi L  z_{w}-2  \pi \alpha/(1-\theta)$
and the bold line shows the behaviour of the counting function $Z_{L,w}$.}
\label{Zfig}
\end{figure}
Working out the terms in (\ref{fsfourierNLIE1}) and computing
the inverse Fourier transform we get a system of non-linear
integral equations (NLIE) for the counting functions:
\be
\label{fsNLIE}
\left( \begin{array}{c}
Z_{L,u} \\
Z_{L,w} \\
Z_{L,v}
\end{array}
\right)
=
2 \pi
L
\left( \begin{array}{c}
z_{u}  \\
z_{w}  \\
z_{v}
\end{array}
\right)
+
\lim_{\eta \rightarrow 0^{+}}
2 \, \textrm{Im}
\,
\mathbf{G}
*
\left( \begin{array}{c}
Q_{L,u} \\
Q_{L,w} \\
Q_{L,v}
\end{array}
\right)
-
\frac{\pi \, \alpha}{(1-\theta)}
\,
\left( \begin{array}{c}
  1 \\
  2 \\
  1 \\
\end{array}
\right)
\ee
where the bulk quantities are given by:
\be
z_{w}(x)
=
\frac{1}{\pi} \arctan \Big( \tanh(\pi x/8) \Big)
\qquad
z_{u}(x)=z_{v}(x)
=
\frac{1}{2\pi} \arctan \Big(\sqrt{2} \sinh(\pi x/4) \Big).
\ee
We will need their asymptotic behaviour:
\be
\label{fsasymptz}
\lim_{x \rightarrow + \infty} z_{u,v}(x)
\sim
\frac{1}{4} -\frac{1}{\sqrt{2} \pi} \,
\textrm{exp}[-\pi \, x / 4]
\qquad
\lim_{x \rightarrow + \infty} z_{w}(x)
\sim
\frac{1}{4} - \frac{1}{\pi} \,
\textrm{exp}[-\pi \, x / 4].
\ee
A plot of the various terms entering in the 
second equation of (\ref{fsNLIE})
is shown in figure~\ref{Zfig}.
Similar plots hold for the first and third equation.
%
\section{Integral expression for the free energy \label{sectionENERGY}}
In this section we will derive an integral 
expression for the finite-size free energy.
First we have to express the free
energy as the sum of some function evaluated
at the BA roots and then applying formula 
(\ref{fsKEYRESULT}) convert
the finite sum into an integral.
For that purpose notice that for the 
distribution of the BA roots corresponding
to the largest eigenvalue the following
equality is satisfied \cite{D}:
\be
\label{fsexpIPIa}
\prod_{j=1}^{n_{u}} S_{1}(u_{j},\theta)^{1/2}
\prod_{k=1}^{n_{v}} S_{1}(v_{k},\theta)^{1/2}
=
\textrm{exp}[\ii \, \pi \, \alpha]
\ee
which allows to write the finite-size free energy 
(\ref{fsFL},\ref{fsAUTOVALORE}) as:
\be
\label{fslaenergia1}
f(L)
=
-
\frac{1}{L}
\left[
\log 2 + \log \cos (2 \, \pi \, \alpha + s_{2})
+\frac{1}{2} \, s_{1}
\right]
\ee
where the finite sums $s_{1}$
and $s_{2}$ are defined by:
\begin{eqnarray}
\label{fsS1S2}
s_{1}
& = &
\sum_{m=1}^{n_{w}} \log \left( \cos^2 \pi \, \theta + \sin^2 \pi \, \theta \, 
\textrm{coth}^2 \frac{\pi \, \theta \, w_{m}}{2} \right)
\\ \nonumber
s_{2}
& = &
\sum_{m=1}^{n_{w}} \textrm{arctan} 
\left( 
\tan \pi \, \theta \, \textrm{coth} 
\left( 
\pi \, \theta \, w_{m}/2
\right) 
\right).
\end{eqnarray}
We will show in the Appendix that only
$s_{1}$  contributes to the value 
of the central charge.
In the sequel we will manipulate the sum
$s_{1}$ in order to extract the exact value
of the central charge.
The function entering in the sum $s_{1}$
(\ref{fsS1S2}) 
and its derivative are defined by:
\be
\label{fsfuncphi1}
\varphi(w)
\equiv
\log 
\left( 
\cos^2 \pi \, \theta + \sin^2 \pi \, \theta \, 
\textrm{coth}^2 \frac{\pi \, \theta \, w}{2} 
\right)
\qquad
W(w)
\equiv
\frac{d}{dw}
\varphi(w).
\ee
At this point remember that in deriving the key
result (\ref{fsKEYRESULT}) which allows the 
conversion of a finite sum into an integral
we made the fundamental assumption that the
function entering in the sum was analytic in
the region delimited by the contour integral.
And moreover that the BA roots where located
inside that region.
The analyticity requirement is not satisfied
in this particular case because $\varphi(w)$
(\ref{fsfuncphi1}) 
is not analytic at the origin.
To overcome the problem we apply the 
formalism of section \ref{sectionSUMINTEGRAL} 
taking as contour line two boxes, one positioned
on the right and the other on the left side 
of the origin, that contain all the $w_{j}$ roots.
This is illustrated in figure~\ref{ex1ufig}~(b).
Now, besides the border terms at infinity,
there will be also border terms at the origin.
They are studied in the Appendix.
The finite sum $s_{1}$ in integral form becomes:
\begin{eqnarray}
\label{fssommas1A}
s_{1}
& = &
-
\textrm{Vp} 
\int_{-\infty}^{+\infty} \frac{dx}{2 \pi}  \, W(x) \, Z_{L,w}(x)
\\ \nonumber
& + &
2 \, \textrm{Im} \, \textrm{Vp} 
\int_{-\infty}^{+\infty} \frac{dx}{2 \pi} \, W(x) \, \log \Big[ 1+\textrm{exp}[\ii \, Z_{L,w}(x+\ii \eta)] \Big]
+
\textrm{border terms}
\end{eqnarray}
\begin{figure}[!t]
\centerline{\epsfxsize=10cm \epsfbox{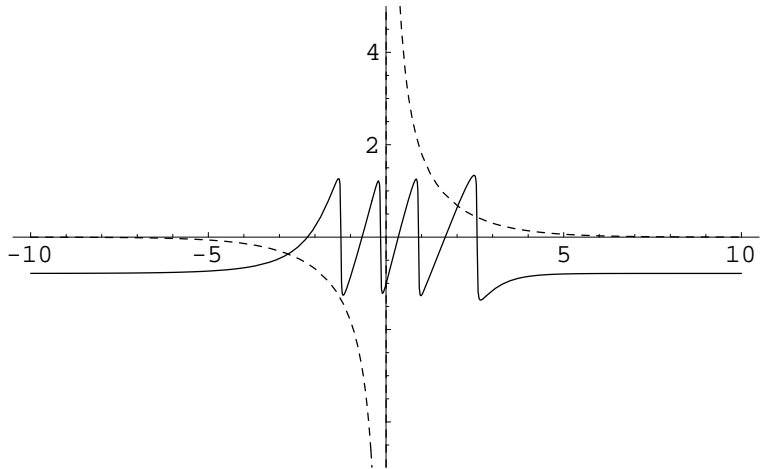}}
\caption{Plot of the function $H_{w}(x)$ (dashed line)
and of the oscillating non-linear term 
$\textrm{Im} \log[1+ \textrm{exp}[\textrm{i} \, Z_{L,w}(x + \textrm{i} \, \eta)]]$.
We refer to a system of size $L=8$, phases $\alpha=\theta=1/4$ 
and parameter $\eta=0.01$.
At the four BA roots $w_{j}$ the non-linear term exhibits a rapid variation.
}
\label{WZfig}
\end{figure}
where Vp denotes the principal value.
The first term on the rhs of (\ref{fssommas1A}) 
can be rewritten substituting the expression 
for the counting function given by 
the non-linear integral equation (\ref{fsNLIE}).
In order to perform the calculations it is convenient
to apply Parseval formula so that $W(x)$, $Z_{L,w}(x)$
and the non-linear term can be replaced by their respective
Fourier transform, and then make use of the expression 
(\ref{fsfourierNLIE1}) for the counting function.  
Performing the calculations in the Fourier space
and then applying again Parseval formula 
via the inverse Fourier transform we get the 
following integral expression for the sum $s_{1}$:
\begin{eqnarray}
\label{fssommas1B}
s_{1}
& = &
-L
\, 
\textrm{Vp}
\int_{-\infty}^{+\infty} dx \, W(x) \, z_{w}(x)
+
\frac{\alpha}{1 - \theta}
\textrm{Vp}
\int_{-\infty}^{+\infty} dx \, W(x)
\\ \nonumber
& &
+
2 \, \textrm{Im} \, \textrm{Vp} \int_{-\infty}^{+\infty} \frac{dx}{2 \pi} H_{u}(x)
\,
\log \Big[ 1+\textrm{exp}[\ii \, Z_{L,u}(x+\ii \eta)] \Big]
\\ \nonumber
& &
-
2 \, \textrm{Im} \, \textrm{Vp} \int_{-\infty}^{+\infty} \frac{dx}{2 \pi} H_{w}(x)
\,
\log \Big[ 1+\textrm{exp}[\ii \, Z_{L,w}(x+\ii \eta)] \Big]
\\ \nonumber
& &
+
2 \, \textrm{Im} \, \textrm{Vp} \int_{-\infty}^{+\infty} \frac{dx}{2 \pi} H_{v}(x)
\,
\log \Big[ 1+\textrm{exp}[\ii \, Z_{L,v}(x+\ii \eta)] \Big]
\\ \nonumber
& &
+
\textrm{border terms}
\end{eqnarray}
where the functions $H_{s}(x)$ are given by:
\be
H_{u}(x)=H_{v}(x) 
=
-\frac{\pi}{\sqrt 2} \frac{\sinh (\pi x/4)}{\cosh (\pi x/2)}
\qquad
H_{w}(x) = \frac{\pi}{2} \textrm{csch} (\pi x/4).
\ee
It is remarkable to notice that the functions $H_{s}(x)$
do not depend on the parameter $\theta$.
The $\theta$ dependence is carried by the logarithmic factors
which contain the counting functions.
Notice that the rhs of (\ref{fssommas1B}) gives an exact expression
for the finite-size sum $s_{1}$ in terms of the finite-size
counting functions $Z_{L,s}$.

Let us examine the various terms that appear in rhs of (\ref{fssommas1B}).
The first term is the dominant one and 
is responsible for the bulk free energy.
The second one comes from the constant part in the NLIE (\ref{fsNLIE}).
It gives a contribution to the finite-size free 
energy density which scales like $1/L$.
Its contribution exactly cancels with a same 
contribution coming from the border terms.
The last three integrals, 
which exhibit a non-linear dependence 
on the counting functions, 
will generate the higher order corrections.
A plot of the function $H_{w}(x)$ and of the non-linear part
$\textrm{Im} \, \log[1+\textrm{exp}[\textrm{i} \, Z_{L,w}(x+\textrm{i} \, \eta)]]$
is shown in figure~\ref{WZfig}.
The non-linear part exhibits rapid jumps at the BA roots $w_{j}$.
Due to the rapid oscillations, it is responsible for the
vanishing of the integral in the limit $L \rightarrow \infty$.
In the next section we are going to extract, from these integrals,
the corrections of order $1/L^2$ which determine
the value of the central charge.

%
\section{The central charge \label{sectionCHARGE}}
In order to compute the central charge we have to manipulate
the NLIE (\ref{fsNLIE}) and the expressions (\ref{fslaenergia1}) 
and (\ref{fssommas1B}) for the free energy.
We make the following change of variable \cite{fsDDV1}:
\be
\label{fscambiodivariabile}
x
=
4(\mu+\log L)/\pi
\ee
which depends on the size of the system,
and introduce the functions:
\be
\begin{tabular}{l l}
$
F_{L,s}(x) 
\equiv 
\textrm{exp}[\ii \, Z_{L,s}(x)]
$
&
\\
$
F_{+,s}(\mu)
\equiv
\lim_{L \rightarrow + \infty}
F_{L,s}(4(\mu + \log L)/\pi)
$
&
$
Q_{+,s}(\mu)
\equiv
2 \, \textrm{Im} \, \log [1+F_{+,s}(\mu+\ii\, \eta)].
$
\end{tabular}
\ee
Making use of the asymptotics (\ref{fsasymptz})
we see that in the limit 
$L \rightarrow + \infty$
the NLIE (\ref{fsNLIE}) reduces to:
\begin{eqnarray}
\label{fsNLIE2}
-\ii \log
\left[ \begin{array}{c}
F_{+,u}(\mu) \\
F_{+,w}(\mu) \\
F_{+,v}(\mu)
\end{array}
\right]
& = &
-
\left( \begin{array}{c}
\sqrt{2} \\
2        \\
\sqrt{2} 
\end{array}
\right)
e^{-\mu}
-
\frac{\pi \, \alpha}{(1-\theta)}
\,
\left( \begin{array}{c}
  1 \\
  2 \\
  1 \\
\end{array}
\right)
\\ \nonumber
& + &
2 \, \textrm{Im}
\,
\int_{-\infty}^{+\infty} d \mu^{\prime} \,
\mathbf{G}(\mu^{\prime}-\mu) \,
\log
\left[ \begin{array}{c}
1+F_{+,u}(\mu^{\prime}+ \ii \, \eta) \\
1+F_{+,w}(\mu^{\prime}+  \ii \, \eta) \\
1+F_{+,v}(\mu^{\prime}+  \ii \, \eta) \\
\end{array}
\right].
\end{eqnarray}
This is a non-linear integral equation for the
unknown functions 
$F_{+,s}(\mu)$.
The dependence on the size of the system $L$
has dropped.
In particular comparing with (\ref{fsNLIE})
we see that the terms of order $L$ has been
reduced to exponentials.
We show now that performing the change of variable 
(\ref{fscambiodivariabile}) in the expression 
for the free energy it is possible
to extract the correction of order $1/L^2$.
First notice that the functions $H_{s}(x)$ that enter
in the integrals (\ref{fssommas1B}) have the
following asymptotic behaviour:
\be
\lim_{x \rightarrow + \infty} H_{u,v}(x) 
\sim 
-
\frac{\pi}{\sqrt{2}} \,  \textrm{exp}[-\pi \,  x/4]
\qquad
\lim_{x \rightarrow + \infty} H_{w}(x) 
\sim 
\pi \, \textrm{exp}[- \pi \, x/4].
\ee
%
In the new variable $\mu$ the leading terms are:
\be
\lim_{L \rightarrow + \infty} H_{u,v}(4(\mu + \log L)/\pi) 
\sim 
-
\frac{\pi}{\sqrt{2}\, L} \, e^{-\mu}
\qquad
\lim_{L \rightarrow + \infty} H_{w}(4(\mu + \log L)/\pi) 
\sim 
\frac{\pi}{L} \, e^{ - \mu}.
\ee
so that the expression for the free energy 
(\ref{fslaenergia1}, \ref{fssommas1B}) reduces to:
\be
\label{fsscalingenergy}
f(L)
\sim
f(\infty)
+
\frac{C_{+}+C_{-}}{2 \, \pi \, L^2}
+
o(1/L^2)
\ee
where the constant $C_{+}$ is defined in terms
of integrals of the functions $F_{+,s}(\mu)$:
\begin{eqnarray}
\label{fslaenergia2}
C_{+}
\nonumber
& = &
2 \, \textrm{Im} \, \int_{-\infty}^{+\infty} d\mu 
 \, \sqrt{2} e^{-\mu} 
\, \log [1+F_{+,u}(\mu + \ii \, \eta)]
\\ 
& + &
2 \, \textrm{Im} \, \int_{-\infty}^{+\infty} d\mu
 \, 2  e^{-\mu}  
\, \log [1+F_{+,w}(\mu + \ii \, \eta)]
\\ \nonumber
& + &
2 \, \textrm{Im} \, \int_{-\infty}^{+\infty} d\mu
 \, \sqrt{2} e^{-\mu} 
\,  \log [1+F_{+,v}(\mu + \ii \, \eta)].
\end{eqnarray}
Notice that the effect of the change of variable 
(\ref{fscambiodivariabile}) 
on the last three integrals of (\ref{fssommas1B}) is twofold:
It shifts the oscillating terms so that only one of
the two tails of the functions $H_{s}$ will 
contributes to the integral, 
and in the limit $L \rightarrow +\infty$
it extracts from the function $H_{s}(4(\mu+\log L)/\pi)$
the leading term $e^{-\mu}/L$.
In order to compute the contribution to 
the finite-size corrections coming from the other tail
we have to perform the change of variable $x=4(\mu-\log L)/\pi$.
We denote the two contributions by $C_{+}$ and $C_{-}$.

Fortunately, it is not necessary to solve the
non-linear integral equation (\ref{fsNLIE2})
in order to calculate the constant $C_{+}$.
In fact we can make use of the
following formula \cite{fsPZJ1}:
\begin{eqnarray}
\label{fsgeneraltrick}
C_{+} & = & -2 \, \textrm{Re} \sum_{s} \int_{\Gamma_{s}} \frac{d u}{u} \log(1+u)
\\ \nonumber
& &
-\frac{1}{2} \, \sum_{r,s} Q_{+,r}(+\infty) Q_{+,s}(+\infty) \int_{-\infty}^{+\infty} dx \, G_{r,s}(x)
\\ \nonumber
& &
+\frac{1}{2} \,  \sum_{r,s} Q_{+,r}(-\infty) Q_{+,s}(-\infty) \int_{-\infty}^{+\infty} dx \, G_{r,s}(x)
\end{eqnarray}
where $\Gamma_{s}$ is an arbitrary line of integration 
in the complex plane connecting
$F_{+,s}(-\infty+ \ii \, 0^{+})$ to $F_{+,s}(+\infty + \ii \, 0^{+})$ 
and avoiding the logarithmic cut $(-\infty,-1]$.
This is the multicomponent generalization 
of the lemma proved in \cite{fsDDV2}.
For completeness we reproduce the
proof in the Appendix.
The limit quantities that enter in the previous 
formula have the following values:
\be
\begin{tabular}{l l l}
$F_{+,s}(-\infty + \ii \, 0^{+})=0$
&
$Q_{+,s}(-\infty)=0$
&
$s=u,w,v$
\\
$F_{+,s}(+\infty + \ii \, 0^{+})=1$
&
$Q_{+,s}(+\infty)=0$
&
$s=u,v$
\\
$F_{+,w}(+\infty + \ii \, 0^{+})=\textrm{exp}[-2 \, \ii \, \pi \, \alpha]$
&
$Q_{+,w}(+\infty)=-2 \, \pi \, \alpha$.
\end{tabular}
\ee
The integrals involving $G_{r,s}$ can be calculated
with the help of the Fourier transform (\ref{fsmatrixG}).
Summing all the contributions, $C_{+}$ becomes:
\be
C_{+}
=
\Big[
-\frac{\pi^2}{6}
\Big]_{u}
+
\Big[
-\frac{\pi^2}{6}
+\frac{4 \, \pi^2 \, \alpha^2}{2}
\Big]_{w}
+
\Big[
-\frac{\pi^2}{6}
\Big]_{v}
+\frac{4 \, \pi^2 \, \alpha^2}{2} \, \frac{\theta}{1-\theta}
=
-
\frac{\pi^2}{6}
\left(
3-\frac{12 \, \alpha^2}{1-\theta}
\right).
\ee
The three square brackets refer to the
dilogarithm part in formula 
(\ref{fsgeneraltrick}) for the three families
of BA roots, respectively $u$, $w$ and $v$.
The dilogarithms are worked out using the
following formula:
\be
\label{fsDlog1}
-2 \, \textrm{Re} \int_{0}^{e^{i \omega}} \frac{du}{u} \log(1+u)
=
-\frac{\pi^2}{6} + \frac{\omega^{2}}{2}.
\ee
The same result holds for $C_{-}$ 
so that the central charge 
(\ref{caricacentrale}) is given by:
\be
\label{fsCHARGE}
c=3-\frac{12 \, \alpha^2}{1- \theta}
\ee
which is in agreement with the result found
in \cite{fsJK1,fsJZ}.
In particular we see that for the four-coloring
problem \cite{fsJK1,fsKH1,fsKH2}, 
which corresponds to the choice 
$\theta=\alpha=0$ the central charge is $c=3$.
For the double Hamiltonian walk 
($\theta=\alpha=1/2$) \cite{fsJK1,fsKONDEV4} 
the central charge becomes negative $c=-3$.
In general, 
the shift of the value of the central charge
due to the presence of the seam has been
investigated in \cite{fsBCN,fsDF1}.
%
%
\section{The excitations \label{sectionEXCITATION}}
\begin{figure}[!t]
\centerline{\epsfxsize=7cm \epsfbox{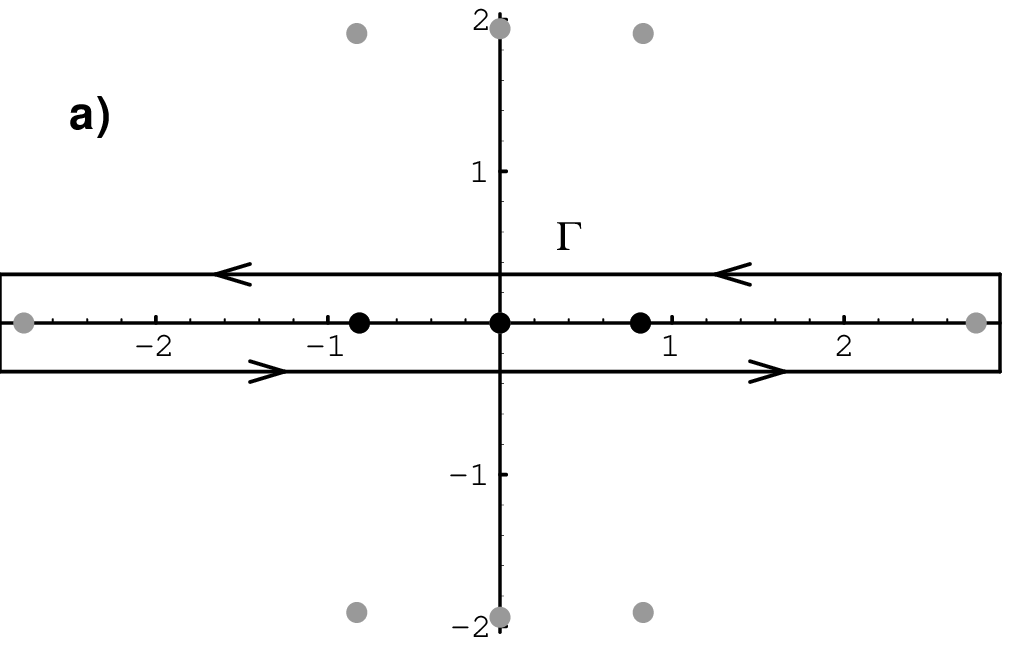} \epsfxsize=7cm \epsfbox{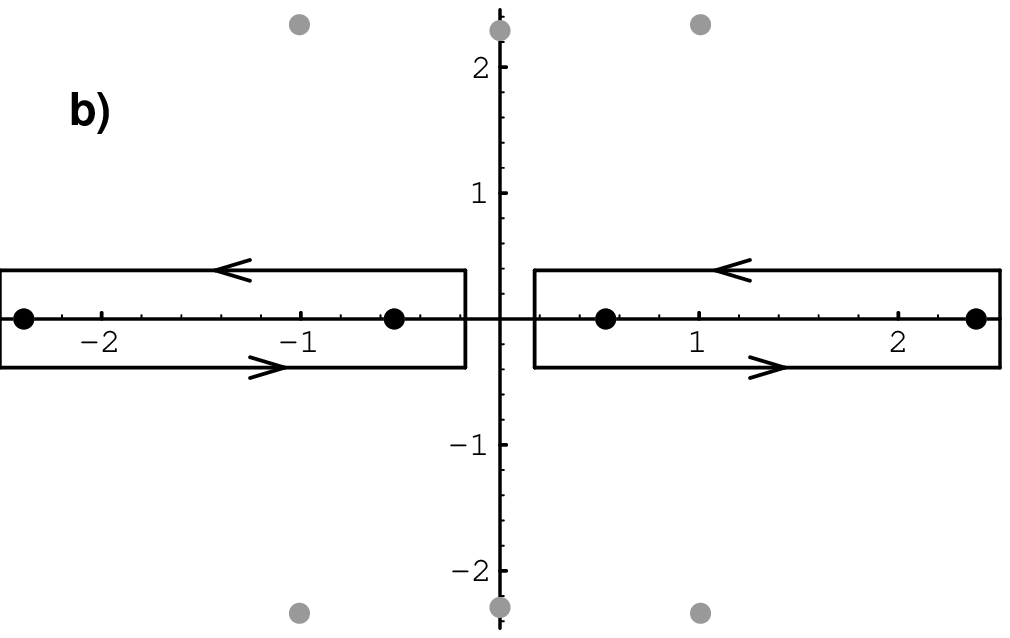}}
\caption{Arrangement of the roots (black circles) 
and of the holes (grey circles) in the complex plane.
a) The $u$-family. b) The $w$-family.
The parameters are:
$L=8$, $n_{v}=n_{w}=8$, $n_{u}=7$, $\alpha=\theta \rightarrow 0$.
The $u$-family has two real holes.
The contour lines which encircles the Bethe
ansatz roots are shown.
a ) Notice that for the $u$-family the contour line
encircles also the two real holes.
b ) In deriving an integral expression for the free
energy we avoid the origin splitting the
contour line into two boxes.
}
\label{ex1ufig}
\end{figure}
The largest eigenvalue of the transfer matrix
lies in the sector $n_{u}=n_{v}=n_{w}=L/2$ and
corresponds to roots distributed on the real 
axis in such a way that the close packing
constraint is satisfied \cite{D}:
\be
Z_{L,s}(\lambda_{s,k+1})
-
Z_{L,s}(\lambda_{s,k})
=
2 \, \pi
\qquad
k=1,\ldots,n_{s}-1
\qquad
s=u,w,v
\ee
Subdominant eigenvalues of the transfer matrix can be                      
constructed by perturbing this distribution \cite{fsBogoliubov}.
One way of doing that (particle-hole excitations)
is to insert real holes $h_{s,k}$ that will break
the close packing of the real roots.
%
%
Another family of excitations (charge excitations)
can be constructed by varying the parameters 
$n_{u}$, $n_{w}$ and $n_{v}$
that enter in the BAE 
(\ref{fsAlgebraicBA1}-\ref{fsAlgebraicBA2}).
A third family of excitations (umklapp processes)
can be constructed shifting the ground state 
distribution of the roots.
To obtain this last excitation we have to replace
in (\ref{fsCONTA}) 
the integer numbers $\textrm{I}_{s,k}$, that
characterize the distribution of the roots
corresponding to the largest eigenvalue, by 
$\textrm{I}_{s,k} \rightarrow \textrm{I}_{s,k} \pm 1$
for $k=1,\ldots,n_{s}$.

In the following we are going to compute analytically,
with the help of formula (\ref{fsDELTA}),
the scaling dimension $\Delta_{i}$ associated with the excitation
in which $n_{u}$ is decreased by one.
We checked numerically that for the distribution of
the BA roots associated with this particular excitation
relation (\ref{fsexpIPIa}) is satisfied so that
the finite-size energy associated with 
the excited eigenvalue $\Lambda_{i}(L)$ is given by:
\be
\log \Lambda_{i}(L)
=
\log 2 + \log \cos (2 \, \pi \, \alpha + s_{2})
+\frac{1}{2} \, s_{1}
\ee
where the sums $s_{1}$ and $s_{2}$ are defined by 
(\ref{fsS1S2}).
We show in the Appendix that also in this case
the contribution to the value of the scaling
dimension comes exclusively from the sum $s_{1}$.
Thus, the machinery developed in section \ref{sectionCHARGE} 
for the calculation of the central charge 
can be applied, with small changes,
also in this case.
Solving numerically the BA equations for
the excited state we see that two real holes
appear for the $u$-family.
This is shown in figure~\ref{ex1ufig}~(a).
There are no real holes for the other
two families of roots.
To denote the position of the real $u$-holes we introduce
the positive real numbers $h_{1}$ for the hole located
on the positive real axis and $h_{2}$ for the
hole on the negative axis.
Also in this case, 
following the steps of section \ref{sectionNLIE}, 
we can write a non-linear integral equation 
for the counting function.
For that notice that the sums, 
entering in the counting functions (\ref{fscountfunc}), 
run over the BA roots.
On the other hand the contour integral
is a box centered around the origin and
extending from $-\infty$ to $+\infty$,
so that it will encircle also the two holes
located at the border of the $u$-roots (see figure~\ref{ex1ufig}~(a)).
At the end their contribution must be subtracted.
The analog of Eq.~(\ref{fsNLIE1}) becomes: 
\begin{eqnarray}
\label{fsNLIE1excitation}
\nonumber
\left( \begin{array}{c}
Z_{L,u}(x) \\
Z_{L,w}(x) \\
Z_{L,v}(x)
\end{array}
\right)
& = &
-
\frac{L}{2} \Phi_{1}
\left( \begin{array}{c}
1 \\
0 \\
1
\end{array}
\right)
+
\mathbf{\Phi}
*
\left( \begin{array}{c}
Z_{L,u} \\
Z_{L,w} \\
Z_{L,v}
\end{array}
\right)
-
\left( \begin{array}{c}
0 \\
2 \, \pi \, \alpha \\
0
\end{array}
\right)
\\
& + &
\left( \begin{array}{c}
\phi_{2}(h_{1}-x)+\phi_{2}(-h_{2}-x) \\
-\phi_{1}(h_{1}-x)-\phi_{1}(-h_{2}-x) \\
0
\end{array}
\right)
\\ \nonumber
& - & 
2 \, \textrm{Im}
\,
\int_{-\infty}^{+\infty} d x^{\prime} \,
\mathbf{G}(x^{\prime}-x) \,
\log
\left[ \begin{array}{c}
1-\textrm{exp}[\ii \, Z_{L,u}(x^{\prime}+ \ii \, \eta)] \\
1+\textrm{exp}[\ii \, Z_{L,w}(x^{\prime}+ \ii \, \eta)] \\
1+\textrm{exp}[\ii \, Z_{L,v}(x^{\prime}+ \ii \, \eta)] \\
\end{array}
\right].
\end{eqnarray}
Notice that in the non-linear term of (\ref{fsNLIE1excitation})
there is a minus sign in front of 
$\textrm{exp}[\ii \, Z_{L,u}]$.
This is because for the excited state we are
investigating the counting function $Z_{L,u}$ evaluated
at the BA roots and holes is an even multiple of $\pi$.
With this change of sign it turns out that the function
$(1-\textrm{exp}[\ii \, Z_{L,u}])$ has a zero of order
one at the BA roots and holes, a necessary condition
to apply the machinery developed 
in section \ref{sectionSUMINTEGRAL}.
Repeating the calculations of section \ref{sectionNLIE}
we can write the analogue of (\ref{fsNLIE}) for the particular
excited state we are investigating:
\begin{eqnarray}
\label{fsNLIEexcitation}
\nonumber
\left( \begin{array}{c}
Z_{L,u}(x) \\
Z_{L,w}(x) \\
Z_{L,v}(x)
\end{array}
\right)
& = &
2 \, \pi \, L
\left( \begin{array}{c}
z_{u} \\
z_{w} \\
z_{v}
\end{array}
\right)
-
\frac{\pi \, \alpha}{(1-\theta)}
\,
\left( \begin{array}{c}
  1 \\
  2 \\
  1 \\
\end{array}
\right)
+
\left( \begin{array}{c}
S_{u}(x+h_{2})+S_{u}(x-h_{1}) \\
S_{w}(x+h_{2})+S_{w}(x-h_{1}) \\
S_{v}(x+h_{2})+S_{v}(x-h_{1})
\end{array}
\right)
\\ 
& &
+
2 \, \textrm{Im}
\int_{-\infty}^{+\infty} dx^{\prime} \, \mathbf{G}(x^{\prime}-x)
\log
\left[ \begin{array}{c}
1-\textrm{exp}[\ii \, Z_{L,u}(x^{\prime}+\ii\eta)] \\
1+\textrm{exp}[\ii \, Z_{L,w}(x^{\prime}+i\eta)]   \\
1+\textrm{exp}[\ii \, Z_{L,v}(x^{\prime}+i\eta)]
\end{array}
\right]
\end{eqnarray} 
where the vector $\mathbf{S}$ 
can be expressed in terms
of the inverse Fourier transform of the matrix 
$\mathbf{\tilde{G}}(p,\theta)$ (\ref{fsmatrixG}):
\be
\label{fsfourierSG}
\left( \begin{array}{c}
S_{u}(x) \\
S_{w}(x) \\
S_{v}(x)
\end{array}
\right)
=
\int_{-\infty}^{+\infty} dp \, \frac{\sin xp}{p}
\,
\left( \begin{array}{c}
\tilde{G}_{11} \\
\tilde{G}_{21} \\
\tilde{G}_{31}
\end{array}
\right).
\ee
In particular it turns out that the derivative of 
$\mathbf{S}$ is related to $\mathbf{G}$:
\be
\frac{d}{dx}
\left( \begin{array}{c}
S_{u}(x) \\
S_{w}(x) \\
S_{v}(x)
\end{array}
\right)
=
2 \pi
\left( \begin{array}{c}
G_{11}(x) \\
G_{21}(x) \\
G_{31}(x)
\end{array}
\right).
\ee
From expression (\ref{fsfourierSG}) and (\ref{fsmatrixG}) 
we read the limit values of $\mathbf{S}$:
\be
\lim_{x \rightarrow \pm \infty}
\left( \begin{array}{c}
S_{u}(x) \\
S_{w}(x) \\
S_{v}(x)
\end{array}
\right)
=
\pm
\pi
\left( \begin{array}{c}
(1-4 \, \theta)/(4 - 4 \, \theta) \\
1/(-2 + 2 \, \theta)              \\
1/(-4+4 \, \theta)                \\
\end{array}
\right).
\ee
Performing the change of variable (\ref{fscambiodivariabile})
and taking the limit $L \rightarrow + \infty$,
equation (\ref{fsNLIEexcitation}) reduces to:
\begin{eqnarray}
\label{fsNLIE3}
-\ii \log
\left[ \begin{array}{c}
F_{+,u}(\mu) \\
F_{+,w}(\mu) \\
F_{+,v}(\mu)
\end{array}
\right]
& = &
-
\left( \begin{array}{c}
\sqrt{2} \\
2 \\
\sqrt{2}
\end{array}
\right)
e^{-\mu}
-
\frac{\pi \, \alpha}{(1-\theta)}
\,
\left( \begin{array}{c}
  1 \\
  2 \\
  1 \\
\end{array}
\right)
\\ \nonumber
& &
+
\pi
\left( \begin{array}{c}
(1-4 \, \theta)/(4 - 4 \, \theta) \\
1/(-2 + 2 \, \theta)              \\
1/(-4+4 \, \theta)                \\
\end{array}
\right)
+
\left( \begin{array}{c}
S_{u}(\mu-\mu_{h_{1}}) \\
S_{w}(\mu-\mu_{h_{1}}) \\
S_{v}(\mu-\mu_{h_{1}})
\end{array}
\right)
\\ \nonumber
& &
+ 2 \, \textrm{Im}
\,
\int_{-\infty}^{+\infty} d \mu^{\prime} \,
\mathbf{G}(\mu^{\prime}-\mu) \,
\log
\left[ \begin{array}{c}
1-F_{+,u}(\mu^{\prime} + \ii \eta) \\
1+F_{+,w}(\mu^{\prime} + \ii \eta) \\
1+F_{+,v}(\mu^{\prime} + \ii \eta) \\
\end{array}
\right]
\end{eqnarray}
which is the analogue of (\ref{fsNLIE2})
for the excited state.
We will need the following limit values:
\be
\begin{tabular}{l l}
$F_{+,s}(-\infty +\ii \, 0) = 0$ 
& 
$Q_{+,s}(-\infty) = 0$
\qquad
$s=u,w,v$ 
\\
$F_{+,u}(+\infty) = \textrm{exp}[\ii \, \pi (1-2 \, \theta)]$
&
$Q_{+,u}(+\infty)=-2 \, \pi \, \theta $
\\
$F_{+,w}(+\infty)=\textrm{exp}[\ii \, \pi (-1-2 \, \alpha + \theta)]$
&
$Q_{+,w}(+\infty)=\pi (-1-2 \, \alpha + \theta)$
\\
$F_{+,v}(+\infty)=1$
&
$Q_{+,v}(+\infty)=0$.
\end{tabular}
\label{fslvexcitation1}
\ee
The excitation we are studying modifies also the
integral expression for the finite-size free energy.
In particular in the integral expression 
for the sum $s_{1}$ (\ref{fssommas1B}) 
the following extra term will appear:
\be
\int_{-\infty}^{+\infty} \frac{dx}{2 \, \pi}
W(x) \, (S_{w}(x+h_{2})+S_{w}(x-h_{1})).
\ee
In the following we will compute the 
contribution to the finite-size correction 
coming from the hole located at position $h_{1}$.
Similar calculations hold for the hole 
positioned at $-h_{2}$.
Working out the integral we obtain:
\be
\int_{-\infty}^{+\infty} \frac{dx}{2 \, \pi}
W(x) \, S_{w}(x-h_{1})
=
\log 
\left(
\frac{\textrm{cosh} (h_{1} \, \pi /4) + 1/\sqrt{2}}{\textrm{cosh} (h_{1} \, \pi /4)-1/\sqrt{2}}
\right)
\ee
which has the following asymptotic behaviour:
\be
\lim_{h_{1} \rightarrow +\infty}
\log 
\left(
\frac{\cosh (h_{1} \, \pi /4) + 1/\sqrt{2}}{\cosh (h_{1} \, \pi /4) - 1/\sqrt{2}}
\right)
\sim
2 \, \sqrt{2} \, \textrm{exp}[-h_{1} \, \pi/4].
\ee
Following the steps of section \ref{sectionCHARGE} we see that, 
in the limit $L \rightarrow +\infty$,
the finite-size energy of the excited state reduces to:
\be
\label{fsscalingenergyEX}  
\log \Lambda_{i}(L)
=
\textrm{bulk}
+
\frac{C_{+}+C_{-}}{2 \, \pi \, L}
+o(1/L)
\ee 
where the constant $C_{+}$ is now given by:
\begin{eqnarray}
\label{fsCplusexcited}
C_{+}
& = &
- 2 \, \pi \, \sqrt{2} \, \textrm{exp}[- \mu_{h_1}]
\\ \nonumber
& &
-2 \, \textrm{Im} \, \int_{-\infty}^{+\infty} d\mu 
\, \sqrt{2} e^{-\mu} 
\, \log [1-F_{+,u}(\mu + \ii \, \eta)]
\\ \nonumber
& &
-2 \, \textrm{Im} \, \int_{-\infty}^{+\infty} d\mu
 \, 2  e^{-\mu}  
\, \log [1+F_{+,w}(\mu + \ii \, \eta)]
\\ \nonumber
& &
-2 \, \textrm{Im} \, \int_{-\infty}^{+\infty} d\mu
 \, \sqrt{2} e^{-\mu} 
\,  \log [1+F_{+,w}(\mu + \ii \, \eta)].
\end{eqnarray}
The bulk term in (\ref{fsscalingenergyEX}) 
coincides with the bulk term of the 
ground state eigenvalue (\ref{fsscalingenergy}).
The dominant finite-size correction 
has a different structure.
In fact comparing (\ref{fsCplusexcited}) with 
(\ref{fslaenergia2}) we see 
that there is an extra contribution
$-2 \, \pi \, \sqrt 2 \, \textrm{exp}[-\mu_{h_1}]$
due to the presence of the
hole positioned at $\mu_{h_1}$
and there is a minus sign in front of $F_{+,u}$.
A similar expression can be derived for $C_{-}$
performing the change of variable
$x=4(\mu-\log L)/\pi$ which selects the hole
located at $-\mu_{h_2}$.
Now we are going to rewrite 
(\ref{fsCplusexcited}) 
in a fashion that allows the application of formula 
(\ref{fsgeneraltrick}).
For that notice that the hole located at $\mu_{h_1}$
solves the first equation of the system (\ref{fsNLIE3}) 
for some integer number
$\textrm{I}_{h_1}$:
\begin{eqnarray}
\label{fsNLIE4}
2 \, \pi \, \textrm{I}_{h_1}
& = &
-\sqrt{2} \, \textrm{exp}[-\mu_{h_1}]
-\frac{\pi \, \alpha}{(1-\theta)}
+\frac{\pi (1 - 4 \, \theta)}{4 (1 - \theta)}
+ S_{u}(0)
\\ \nonumber
& + &
2 \, \textrm{Im} 
\int_{-\infty}^{+\infty}
\frac{d\mu}{2 \, \pi}
\frac{d}{d \mu}
S_{u}(\mu - \mu_{h_1})
\log[1-F_{+,u}(\mu+\ii \, \eta)]
\\ \nonumber
& + &
2 \, \textrm{Im} 
\int_{-\infty}^{+\infty}
\frac{d\mu}{2 \, \pi}
\frac{d}{d \mu}
S_{w}(\mu - \mu_{h_1})
\log[1+F_{+,w}(\mu+\ii \, \eta)]
\\ \nonumber
& + &
2 \, \textrm{Im} 
\int_{-\infty}^{+\infty}
\frac{d\mu}{2 \, \pi}
\frac{d}{d \mu}
S_{v}(\mu - \mu_{h_1})
\log[1+F_{+,v}(\mu+\ii \, \eta)]
\end{eqnarray}
where according to (\ref{fsfourierSG}) $S_{u}(0)=0$. 
Now, substituting the expression for the exponential,
given by relation (\ref{fsNLIE4}) into the rhs
of (\ref{fsCplusexcited}) we get the following 
expression for $C_{+}$:
\begin{eqnarray}
C_{+}
& = & 
4 \, \pi^2 \textrm{I}_{h_1}
+ \frac{2 \, \pi^2 \, \alpha}{1-\theta}
- \frac{\pi^2  (1-4 \, \theta)}{2(1 - \theta)}
\\ \nonumber
& - &
2 \, \textrm{Im} \, \int_{-\infty}^{+\infty} d\mu 
 \, 
\Big[
\sqrt{2} e^{-\mu}
+\frac{d}{d \mu} S_{u}(\mu -\mu_{h_1})
\Big]
\, \log [1-F_{+,u}(\mu + \ii \, \eta)]
\\ \nonumber
& - &
2 \, \textrm{Im} \, \int_{-\infty}^{+\infty} d\mu
 \, 
\Big[
2  e^{-\mu} 
+\frac{d}{d \mu} S_{w}(\mu -\mu_{h_1})
\Big]
\, \log [1+F_{+,w}(\mu + \ii \, \eta)]
\\ \nonumber
& - &
2 \, \textrm{Im} \, \int_{-\infty}^{+\infty} d\mu
 \, 
\Big[
\sqrt{2} e^{-\mu} 
+\frac{d}{d \mu} S_{v}(\mu -\mu_{h_1})
\Big]
\,
\log [1+F_{+,v}(\mu + \ii \, \eta)].
\end{eqnarray}
Notice that the expression in square
brackets is the derivative of the first
four terms on the rhs of (\ref{fsNLIE3}) 
so that we are in the position to apply
formula (\ref{fsgeneraltrick}).
Showing all the terms:
\begin{eqnarray}
\label{fsCplusEx}
C_{+}
& = &
4 \, \pi^2 \textrm{I}_{h_{1}}
+ \frac{2 \, \pi^2 \, \alpha}{1-\theta}
- \frac{\pi^2  (1-4 \, \theta)}{2(1 - \theta)}
\\ \nonumber
& - &
\pi^2
\Big(
\Big[
\frac{1}{3}
+
\frac{(1-2 \, \theta)^2}{2}
-
(1-2 \, \theta)
\Big]_{u}
+
\Big[
-
\frac{1}{6}
+
\frac{(-1-2 \, \alpha + \theta)^2}{2}
\Big]_{w}
+
\Big[
-\frac{1}{6}
\Big]_{v}
+
q
\Big).
\end{eqnarray}
For $C_{-}$ we obtain the same expression with 
$\textrm{I}_{h_1}$ replaced by $\textrm{I}_{h_2}$.
The three square brackets refer to the dilogarithm
part in formula (\ref{fsgeneraltrick}) for the three
family of BA roots, respectively $u$, $w$ and $v$.
The dilogarithms are worked out using formula 
(\ref{fsDlog1}) and the limit values
(\ref{fslvexcitation1}).
In order to compute the contribution coming
from the $u$-term we make use of the formula:
\begin{eqnarray}
\nonumber
-2 \, \textrm{Re} \int_{0}^{e^{i \omega}} \frac{du}{u} \log(1-u)
& = &
\frac{\pi^2}{3} + \frac{\omega^2}{2} -\pi \, \omega.
\end{eqnarray} 
The term $q$ in the expression (\ref{fsCplusEx}) gives the
contribution coming from the 
non-dilogarithm part of formula (\ref{fsgeneraltrick}).
Working out we find:
\be
q
=
\frac
{
\theta(3+4 \, \alpha^2 + 8 \, \alpha - 4 \, \alpha \, \theta - 5 \, \theta + 5 \, \theta^2)
}
{
2(1-\theta)
}
\ee
We have all the ingredients to apply formula (\ref{fsDELTA}).
Substituting the values of $C_{-}$ and $C_{+}$ in
(\ref{fsscalingenergyEX}) and reminding that the central charge
is given by (\ref{fsCHARGE}) after some algebra we get 
simple expressions
for the scaling dimensions:
\be
\label{fsSCALINGDIMENSION}
\Delta_{+}
=
-
\textrm{I}_{h_1}
+
\frac{1-\theta}{4}
\qquad
\Delta_{-}
=
-
\textrm{I}_{h_2}
+
\frac{1-\theta}{4}
\qquad
\Delta
\equiv
\Delta_{+}+\Delta_{-}
=
\frac{1-\theta}{2}
\ee
where the $\alpha$ dependence has dropped.
From numerical calculation we see that the
excitation under investigation corresponds to
the choice $\textrm{I}_{h_1}=\textrm{I}_{h_2}=0$.

From this result we can compute the scaling dimension
corresponding to one black and one grey string propagating
between two points on the lattice \cite{fsJK1}.
Call $G_{1}(r)$ the two string correlation function 
associated with defects configurations where a single oriented
black loop segment and a single oriented grey loop segment
propagate from a vertex located at the origin to a vertex
positioned $r$ rows above it.
If we denote by $\phi$ the operator which creates the defect,
the correlation function, in the transfer matrix formalism
is given by:
\begin{eqnarray}
G_{1}(r)
& = &
\langle \phi(r) \phi(0) \rangle 
=
\lim_{N \rightarrow \infty}
\frac{1}{Z} \, 
\textrm{Tr} \,\mathbf{T}^{N-r}_{0,0,0} 
\, \mathbf{\phi}(r) \, \mathbf{T}^r_{-1,0,0} \, \mathbf{\phi}(0)
\\
& = & \nonumber
\sum_{n} e^{-(E_{n}(L)-E_{0}(L)) r }
\langle 0 | \phi(r) | n \rangle  \langle n | \phi(0) | 0  \rangle
\end{eqnarray}
where $\mathbf{T}_{0,0,0}$ is the block of the transfer matrix corresponding
to the ground state sector and $\mathbf{T}_{-1,0,0}$ is the block 
corresponding to the excited sector in which the number of $u$
particles is decreased by one.
Notice that the sum runs over the eigenvalues 
$e^{-E_{n}(L)}$ of $\mathbf{T}_{-1,0,0}$ and $e^{-E_{0}(L)}$
is the largest eigenvalue of the transfer matrix.
In the limit $r \gg 1$ the correlation
exhibits, in the cylindric geometry, an exponential decay:
\be
G_{1}(r) \sim e^{-(E_{1}-E_{0})r}
\ee
where $e^{-E_{1}(L)}$ is the largest 
eigenvalue of the matrix $\mathbf{T}_{-1,0,0}$.
In the bulk ($L \rightarrow \infty$) the
correlator exhibits a power law behaviour:
\be
G_{1}(r) \sim \frac{1}{r^{2 x_{1}}}.
\ee
If the model is conformaly invariant
the energy gap and the critical exponent
can be related via the formula 
(see also formula (\ref{fsDELTA})):
\be
x_{1}
=
\lim_{L \rightarrow \infty}
\frac{L}{2 \, \pi}
\,
(
E_{1}(L)-E_{0}(L)
).
\ee
%
%
It is important to notice that since $\mathbf{T}_{-1,0,0}^r$
generates the configurations in which a black
and a gray string run from the first to the
$r\textrm{th}$ row of the cylinder,
black and grey winding loops 
are forbidden inside this region so that there is
no need to introduce the seam in order to assign 
them the right fugacity \cite{fsKONDEV5,fsBatchelor1}.
Moreover an eventual seam would associate a phase, 
to loops spiralling around the cylinder, 
proportional to the winding number.
Performing the calculation that lead to (\ref{fsSCALINGDIMENSION})
with $\alpha=\theta$ for the largest eigenvalue
and $\alpha=0$ for the subdominant one we get for
the exponent governing the two string correlation function:
\be
\label{fsEXPONENT}
x_{1}=\frac{1-\theta}{2}-\frac{\theta^2}{1-\theta}
\ee
in agreement with the result found in \cite{fsJK1} 
via the field theory.
Notice that the second term in the rhs of 
(\ref{fsEXPONENT}) is 1/12 of the shift of
the central charge 
(see the formula (\ref{fsCHARGE}) for the central charge).
\section{Numerics \label{sectionNUMERICS}}
\begin{figure}[!t]
\centerline{\epsfxsize=7cm \epsfbox{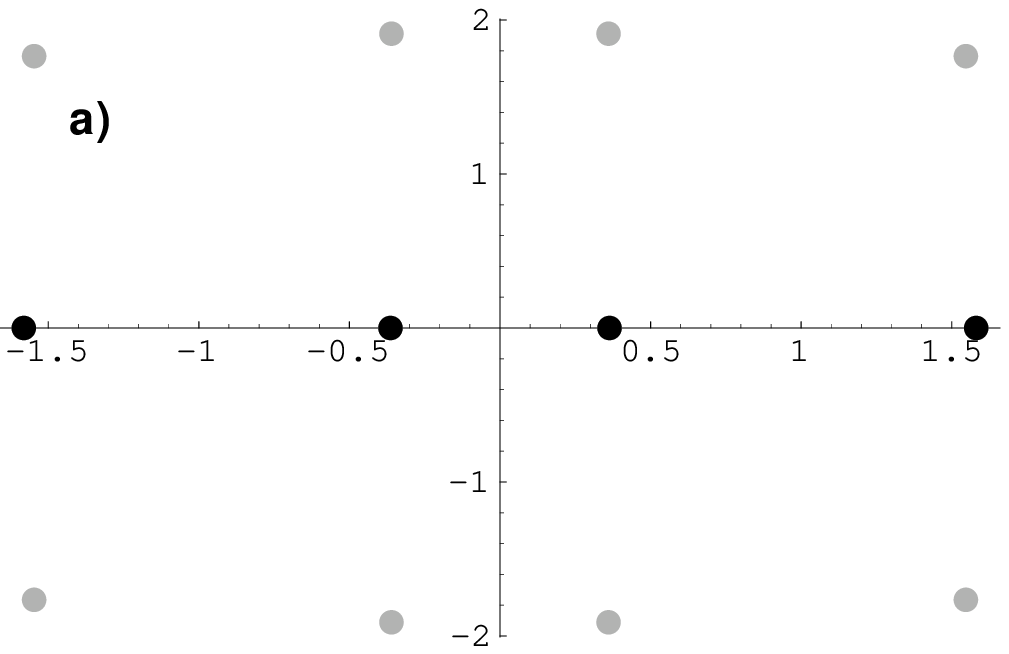} \epsfxsize=7cm \epsfbox{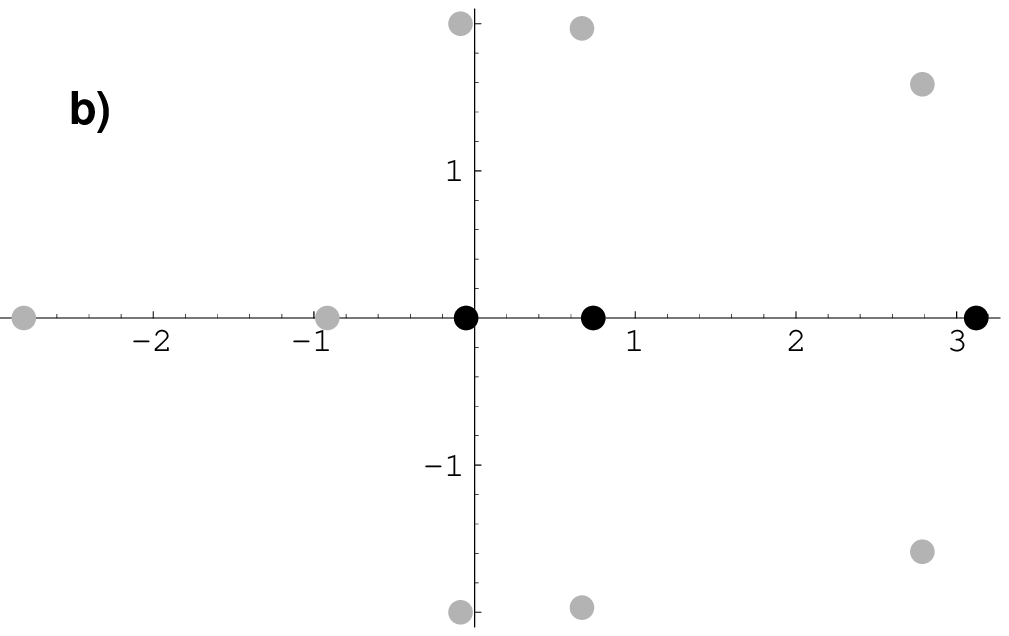}}
\caption{
Arrangement of the roots (black circles) 
and the holes (grey circle) in the complex
plane for the $u$-family.
a) Arrangement corresponding to the ground state.
The parameters are: $L=8$, $n_{u}=n_{v}=n_{w}=4$, $\alpha=\theta=0$.
Notice that the ground state possesses complex holes.
b) Arrangement corresponding to the excitation labelled
by $\mathbf{m}=(-1,0,0)$ and $\mathbf{e}=(1,0,0)$.
Notice that the charge excitation 
$\mathbf{m}=(-1,0,0)$ generates two real holes 
at the border of the Fermi sea (see figure~\ref{ex1ufig} (a)).
The umklapp process $\mathbf{e}=(1,0,0)$ shifts
all the root distribution to the right.
}
\label{umklappfig}
\end{figure}
In the previous section we have shown in detail
how to compute analytically the scaling dimension
associated with the excitation in which the 
parameter $n_{u}$ is decreased by one.
In general, an excitation will be characterized
by a particular arrangement of holes and roots
in the complex plane.
If the arrangement is known the corresponding
scaling dimension can be calculated going 
through the machinery that we have described.
In this section we are going to study numerically
a larger set of excitations.

To classify the excitations in a compact 
manner we introduce three vectors 
$\mathbf{e}=(e_{u},e_{w},e_{v})$,
$\mathbf{m}=(m_{u},m_{w},m_{v})$,
$\mathbf{I^{\pm}}=(\textrm{I}^{\pm}_{u},\textrm{I}^{\pm}_{w},\textrm{I}^{\pm}_{v})$.
The components of $\mathbf{m}$ are integer numbers
which count the change of the ``number of particles''
in the excited state with respect to the ground state.
The components of $\mathbf{e}$ count the total number 
of transitions of the ``particles'' from one side to the
other of the three Fermi seas.
The non-negative integers $\textrm{I}^{\pm}_{s}$
describe excitations in the vicinity of the Fermi surfaces.
They count the number of holes that are generated by moving the roots,
positioned at the Fermi surfaces, to the next vacant positions.
As example, figure \ref{umklappfig} 
shows the arrangement of holes and roots 
corresponding to the excitation labeled by
$\mathbf{m}=(-1,0,0)$, $\mathbf{e}=(1,0,0)$
and $\mathbf{I}^{\pm}=(0,0,0)$ 
for a system of size $L=8$.



Table \ref{fsnumerics1} shows a set of charge excitations
labelled by the vector $\mathbf{m}$.
The scaling dimensions have been estimated by solving
numerically the BA equations for systems of increasing size
and making use of formula (\ref{fsDELTA}).
Notice that in applying formula (\ref{fsDELTA}) the
largest and the subdominant eigenvalues were computed
keeping the value of $\alpha$ constant and not varying it 
as in the derivation of (\ref{fsEXPONENT}).
From the data we see that the scaling dimensions 
do not exhibit a dependence on $\alpha$.
The data for $\mathbf{m}=(-1,0,-1)$ and $\mathbf{m}=(-1,0,+1)$
merit a special comment.
Solving numerically the BA equations for a generic value 
of $\theta$ and setting $\alpha=0$ we see that two particles
at the border of the $w$-family go to infinity.
Specifying the value $w_{l}=\pm \infty$ in (\ref{fsAlgebraicBA2})
and observing that for the function (\ref{fsrational})
$\lim_{x \rightarrow \pm \infty} S_{c}(x,\theta)=-\textrm{exp} [\mp \textrm{i} \,  \pi \, \theta \, c]$
it follows that (\ref{fsAlgebraicBA2}) is satisfied only for $\alpha=0$.
Thus the excitations $\mathbf{m}=(-1,0,-1)$ and $\mathbf{m}=(-1,0,+1)$
we have founded only for $\alpha=0$.

Table \ref{fsnumerics2} shows the scaling dimensions 
for a set of excitations constructed perturbing the 
ground state by a combination of charge excitations
labelled by $\mathbf{m}$ with umklapp processes
labelled by $\mathbf{e}$.
We have checked that for this set of excitations 
in which the umklapp is involved the scaling dimensions
exhibit a dependence on $\alpha$.

Table \ref{fsnumerics3} shows the scaling dimensions
for a set of excitations constructed combining charge
excitations $\mathbf{m}$ with particle-hole excitations
labelled by $\mathbf{I}^{\pm}$.
In order to construct the particle-hole excitations
we insert a hole at the very end of the Fermi sea.
Notice that there are three Fermi seas, one for each
type of particles, 
and each Fermi sea has two surfaces that can be perturbed
independently.
We distinguish among the two surfaces using an 
upper index for the vector $\mathbf{I}^{\pm}$.
From the data we see that the scaling dimensions
for this class of excitations do not depend on
the phase $\alpha$.
Notice that the scaling dimension for the excitation
labelled by $\mathbf{m}=(-1,0,0)$ and 
$\mathbf{I}^{+}=(1,0,0)$ can be calculated analytically
setting in formula (\ref{fsSCALINGDIMENSION}) 
$\textrm{I}_{h_{1}}=-1$ and $\textrm{I}_{h_{2}}=0$

The set of excitations shown in table \ref{fsnumerics1}
and \ref{fsnumerics2} does not cover all the possibilities.
In fact we did not succeed in finding the excitation
labelled by $\mathbf{e}=(+1,+1,0)$ and the one labelled by
$\mathbf{m}=(+1,-1,0)$.
The numerical data, of the low-lying excitations we could construct, 
suggest that the scaling dimensions $\Delta$,
are described by the formula:
\be
\label{fsFORMULAdimension}
\Delta(\mathbf{m},\mathbf{e},\mathbf{I}^{\pm},\theta)
=
\frac{1-\theta}{4} \, \mathbf{m}^{t} \, \mathbf{C} \, \mathbf{m} 
+\frac{1}{1-\theta} \, \mathbf{e}^{t} \, \mathbf{C}^{-1} \, (\mathbf{e} - \mathbf{e_{0}})
+\sum_{s} 
\textrm{I}^{+}_{s}
+\sum_{s}
\textrm{I}^{-}_{s} 
\ee
where the background charge is given by $\mathbf{e_{0}}=(0, 2 \, \alpha,0)$,
and $\mathbf{C}$ is the Cartan matrix of $s\ell_{4}$:
\begin{table}[!h]
\begin{center}
\begin{tabular}{|c|c|c|c|c|c|c|}
\hline
$ \mathbf{m}$ & $\theta$ & $\alpha$ & $L=100$ & $L=200$ & $L=300$ & $L=\infty$ \\
\hline
\hline
$
\begin{array}{c}
(-1,0,0) \\
(+1,0,0)
\end{array} 
$
& 0.25 & 0 & 0.375018    & 0.375004   & 0.375002 &  $\mathbf{0.375}$ \\
\hline
$
\begin{array}{c}
(-1,0,0) \\
(+1,0,0)
\end{array} 
$
& 0.25 & 0.25 & 0.374946  & 0.374986  & 0.374994 &  $\mathbf{0.375}$ \\
\hline
$
\begin{array}{c}
(0,-1,0) \\
(-1,-1,0)
\end{array} 
$
& 0.25 & 0.25 & 0.375018 & 0.375005 & 0.375002 & $\mathbf{0.375}$ \\
\hline
$
\begin{array}{c}
(-1,0,-1) \\
(-1,0,+1)
\end{array} 
$
& 0.25 & 0 & 0.750368 & 0.750092 & 0.750041 & $\mathbf{0.75}$ \\
\hline
$
\begin{array}{c}
(-2,0,0) \\
(0,-2,0)
\end{array}
$
& 0.25 & 0 & 1.499419 & 1.499854 & 1.499935 & $\mathbf{1.5}$ \\
\hline
(0,-2,0) & 0.25 & 0.25 & 1.499421 & 1.499854 &  1.499935 & $\mathbf{1.5}$ \\
\hline
\end{tabular}
\caption{A list of charge excitations labelled by the vector $\mathbf{m}$.
We group in the same box excitations that have exactly the same
excited eigenvalue.
The scaling dimensions are estimated by solving numerically the
BA equations for systems of increasing size $L$ and making use
of formula (\ref{fsDELTA}).
Notice that the value of the scaling 
dimension does not depend on $\alpha$.
The last column shows the dimension predicted by formula 
(\ref{fsFORMULAdimension}).} \label{fsnumerics1}
\end{center}
\end{table}
\vspace{-0.5cm}
\begin{table}[h!]
\begin{center}
\begin{tabular}{|c|c|c|c|c|c|c|}
\hline
$\mathbf{m}$,  $\mathbf{I}^{+}$ & $\theta$ & $\alpha$ & $L=100$ & $L=200$ & $L=300$ & $L=\infty$ \\
\hline
\hline
$
\begin{array}{c}
\mathbf{m}=(-1,0,0)   \\
\mathbf{I}^{+}=(1,0,0)
\end{array} 
$
& 0.25 & 0 & 1.373801 & 1.374700 & 1.374867  &  $\mathbf{1.375}$ \\
\hline
$
\begin{array}{c}
\mathbf{m}=(-2,-2,-2)   \\
\mathbf{I}^{+}=(1,0,0)
\end{array} 
$
& 0.25 & 0 & 2.495597 & 2.498891 &  2.499506 &  $\mathbf{2.5}$ \\
\hline
$
\begin{array}{c}
\mathbf{m}=(-2,-2,-2)   \\
\mathbf{I}^{+}=(1,0,0)
\end{array} 
$
& 0.25 & 0.25 & 2.4966007 & 2.499141 & 2.499617  &  $\mathbf{2.5}$ \\
\hline
$
\begin{array}{c}
\mathbf{m}=(-2,-3,-2)   \\
\mathbf{I}^{+}=(0,1,0)
\end{array} 
$
& 0.25 & 0 & 2.871114 & 2.874015 & 2.874560 &  $\mathbf{2.875}$ \\
\hline
$
\begin{array}{c}
\mathbf{m}=(-2,-3,-2)   \\
\mathbf{I}^{+}=(0,1,0)
\end{array} 
$
& 0.25 & 0.25 & 2.873027 & 2.874490 & 2.874771 & $\mathbf{2.875}$ \\
\hline
\end{tabular}
\caption{A list of excitations constructed combining
the charged excitation labelled by the vector $\mathbf{m}$ with the
particle-hole excitation labeled by $\mathbf{I}^{+}$.
The scaling dimensions are estimated by solving numerically
the BA equations for systems of increasing size $L$ and
making use of formula (\ref{fsDELTA}).
In order to create the particle-hole excitation we insert
a hole between the two particles that are located at the
very end of the Fermi sea.
Notice that for this class of excitations the scaling
dimensions do not depend on the phase $\alpha$.
The last column shows the dimensions predicted by 
formula (\ref{fsFORMULAdimension}).
\label{fsnumerics3}}
\end{center}
\end{table}
\begin{table}[!h]
\begin{center}
\begin{tabular}{|c|c|c|c|c|c|c|}
\hline
$ \mathbf{m},\mathbf{e}$ & $\theta$ & $\alpha$ & L=100 & L=200 & L=300 & $L=\infty$ \\
\hline
\hline
$
\begin{array}{c}
\mathbf{e}=(-1,0,0) \\
\mathbf{m}=(-1,0,0)
\end{array} 
$
& 0.25 & 0 & 1.373677 & 1.374668 & 1.374852 & $\mathbf{1.375}$ \\
\hline
$
\begin{array}{c}
\mathbf{e}=(-2,0,0) \\
\mathbf{m}=(-4,-4,-4)
\end{array} 
$
& 0.25 & 0 &  9.922879 & 9.980360 & 9.991231 & $\mathbf{10}$ \\
\hline
$
\begin{array}{c}
\mathbf{e}=(-1,0,+1) \\
\mathbf{m}=(-2,-2,-2)
\end{array} 
$
& 0.25 & 0 &  2.829248 & 2.832285 & 2.832863 & $\mathbf{2.8\overline{3}}$ \\
\hline
$
\begin{array}{c}
\mathbf{e}=(+1,0,+1) \\
\mathbf{m}=(-2,-2,-2)
\end{array} 
$
& 0.25 & 0 & 4.154368 & 4.163582 & 4.165295 & $\mathbf{4.1\overline{6}}$ \\
\hline
$
\begin{array}{c}
\mathbf{e}=(0,+1,0) \\
\mathbf{m}=(-1,-2,-1)
\end{array} 
$
& 0.25 & 0 & 2.087041 & 2.084259 & 2.083745  & $\mathbf{2.08\overline{3}}$ \\
\hline
$
\begin{array}{c}
\mathbf{e}=(-1,+1,0) \\
\mathbf{m}=(-2,-2,-2)
\end{array} 
$
& 0.25 & 0 & 2.875721 & 2.875168 & 2.875073 & $\mathbf{2.875}$ \\
\hline
$
\begin{array}{c}
\mathbf{e}=(+1,0,0) \\
\mathbf{m}=(-2,-2,-2)
\end{array} 
$
& 0.25 & 0.3 & 2.097194  & 2.099292 & 2.099684 &  $\mathbf{2.1}$ \\
\hline
$
\begin{array}{c}
\mathbf{e}=(-1,0,0) \\
\mathbf{m}=(-2,-2,-2)
\end{array} 
$
& 0.25 & 0.3 & 2.893517 & 2.898372 & 2.899276 &  $\mathbf{2.9}$ \\
\hline
\end{tabular}
\caption{
A list of excitations constructed combining a variation
in the number of ``particles'' labelled by $\mathbf{m}$,
with transitions, from one side to other of the Fermi seas,
labelled by $\mathbf{e}$.
The scaling dimensions are estimated by solving numerically
the BA equations for systems of increasing size $L$
and making use of formula (\ref{fsDELTA}).
Notice that for this set of excitations in which
$\mathbf{e} \neq (0,0,0)$ the scaling dimensions
exhibit a dependence on $\alpha$.
The last column shows the dimensions predicted by formula 
(\ref{fsFORMULAdimension}). \label{fsnumerics2}
}
\end{center}
\end{table}
\be
\mathbf{C}
=
\left(
\begin{array}{ccc}
2  & -1 &  0 \\
-1 &  2 & -1 \\
0  & -1 &  2
\end{array}
\right)
\ee
which confirms the results proposed independently in \cite{fsJZ}.
A formula of this type was derived 
for the first time in \cite{fsIzergin}
for the conformal dimensions of models 
solvable by multicomponent Bethe ansatz.
A rigorous proof of the fact that the scaling 
dimensions associated with the charge excitation 
and the umklapp process are essentially  
inverse of each other has been given
in \cite{fsBogoliubov}.
A similar expression for the conformal 
dimensions, in which $\mathbf{C}$ is replaced
by the Cartan matrix of $s\ell_{3}$,
has been found in \cite{fsReshetikhin}
for a loop model on the hexagonal lattice.
%
%
%
\section{Conclusions}
We have derived a non-linear integral equation 
for the $\textrm{FPL}^2$ model on the critical part of the solvable line.
As an application we have extracted exact expressions for
the central charge and for the critical exponent associated
with one black and one grey string propagating between
two points on the lattice.
We have also studied numerically the low-lying excitations
that can be constructed perturbing the Fermi seas in a more 
general manner (charge, umklapp and particle-hole excitations)
and found that the scaling dimensions can be classified 
in a compact way through the Cartan matrix of $s\ell_{4}$.
The calculations have been done without having the knowledge
of the underlying group structure discovered in \cite{fsJZ}
and constitute a confirmation of their results.
It would be interesting to continue our study by considering
strings excitations that can be generated selecting solutions
of the BA equations in which some of the roots have an
imaginary part different from zero.
%
\appendix
\section{Appendix}
%
%
\subsection{The sums $s_{1}$ and $s_{2}$}
Here we are going to show that 
the value of the central charge,
which gives the dominant finite-size correction
to the free energy density (\ref{fslaenergia1}),
is determined exclusively by the sum $s_{1}$.
The sum $s_{1}$ and $s_{2}$ has been defined as (\ref{fsS1S2}):
\be
s_{1}=\sum_{m=1}^{n_{w}} \varphi_{1}(w_{m})
\qquad
\varphi_{1}(x)
=
\log 
\Big(
\cos^2 \, \pi \, \theta + \sin^2 \, \pi \, \theta
\,
\textrm{coth}^2 \frac{\pi \, \theta \, x}{2}
\Big)
\ee
\be
s_{2}=
\sum_{m=1}^{n_{w}}
\varphi_{2}(w_{m})
\qquad
\varphi_{2}(x)
=
\textrm{arctan} (\textrm{tan} (\pi \, \theta) \, \textrm{coth} (\pi \, \theta \, x/2)).
\ee
The function $\varphi_{1}(x)$ has a singularity at
the origin which, 
in deriving the integral expression for the free energy,
will generate border terms.
In particular it can be shown that:
\be
\lim_{\eta \rightarrow 0^{+}}
\frac{1}{\pi} \textrm{Im} \, \varphi_{1}(x+\ii \, \eta)
\textrm{log}
\Big[
1+\textrm{exp}[\ii Z_{L,w}(x+\ii \, \eta)]
\Big]
\Big|_{0^{+}}^{0^{-}}
=
2 ( \log 2 + \log \cos (Z_{L,w}(0)/2)).
\ee
We checked numerically with fifteen digits of accuracy,
that, for the ground state and the excited state under investigation,
the term depending on $s_{2}$, 
in the expression for the free energy (\ref{fslaenergia1}),
exactly cancels with the previous border term:
\be
\log \cos (2 \, \pi \, \alpha + s_{2}) - \log \cos (Z_{L,w}(0)/2)=0
\ee
so that the finite-size corrections to the free energy density are
determined exclusively by the sum $s_{1}$.
%
%
\subsection{Lemma}
Here, for completeness,  we reproduce, following \cite{fsDDV2},
the prove of formula (\ref{fsgeneraltrick}). 

$\mathbf{Hypothesis}$:
\\
Assume that $F(x)$ satisfies the non-linear integral equation (NLIE):
\be
\label{equatio1}
-\textrm{i} \, \textrm{log} \, F(x) 
= 
\phi(x) + \int_{-\infty}^{+\infty}
dy \, G(x-y) \, Q(y)
\ee
where the non-linear term $Q(x)$ is defined by:
\be
\label{equatio2}
Q(x) \equiv 2 \, \textrm{Im} \, \textrm{log}[1 + F(x+ \textrm{i} \, \eta)]
\ee
assume also that the functions $\phi(x)$ and $G(x)$ are real and that
$G(x)$ is symmetric, integrable and peaked around the origin.

$\mathbf{Thesis}$:
\\
Then the following identity holds:
\begin{eqnarray}
\label{equatio3}
\int_{-\infty}^{+\infty} dx \, \phi^{\prime}(x) \, Q(x)
& = &
-2 \, \textrm{Re} \int_{\Gamma} \frac{du}{u} \, \textrm{log}[1+u]
\\ \nonumber
&  &
-
\frac{1}{2}
\Big[
Q^{2}(+\infty)-Q^{2}(-\infty)
\Big]
\int_{-\infty}^{+\infty} dx \, G(x)
\end{eqnarray}
where $\Gamma$ is a contour in the complex u-plane from
$F(-\infty+\textrm{i} \, 0^{+})$ 
to 
$F(+\infty+\textrm{i} \, 0^{+})$
avoiding the logarithmic cut $(-\infty,-1]$.

$\mathbf{Proof}$:
\\
Deriving (\ref{equatio1}) and substituting the expression
for $\phi^{\prime}(x)$ into the lhs of (\ref{equatio3}) we get:
\begin{eqnarray}
\label{equatio4}
\int_{-\infty}^{+\infty} dx \, \phi^{\prime}(x) \, Q(x)
& = &
-2 \, \textrm{Im} \int_{-\infty}^{+\infty} 
dx \, \textrm{i} \frac{F^{\prime}(x)}{F(x)}
\, \textrm{log}[1+F(x+\textrm{i} \, \eta)]
\\ \nonumber
&  &
-
\int_{-\infty}^{+\infty} dx 
\Big[
\int_{-\infty}^{+\infty} dy \, G^{\prime}(x-y) \, Q(y)
\Big] Q(x)
\end{eqnarray}
The first term on the rhs of (\ref{equatio4}) originates
the dilogarithmic term in the rhs of (\ref{equatio3}).
Let us examine in detail the double integral $\textrm{I}(a,b)$ where $0 < a < b$:
\begin{eqnarray}
\label{equatio5}
\textrm{I}(a,b)
& = &
\int_{-a}^{+a} dx 
\Big[
\int_{-b}^{+b} dy \, G^{\prime}(x-y) \, Q(y)
\Big]
\,
Q(x)
\\ \nonumber
& = &
\int_{-a}^{+a} dx 
\Big[
\int_{-b}^{-a} dy \,  G^{\prime}(x-y) \, Q(y)
+
\int_{+a}^{+b} dy \,  G^{\prime}(x-y) \, Q(y)
\Big]
\,
Q(x)
\end{eqnarray}
where we have used the fact that $G^{\prime}(x)$ is an odd function.
Integrating by parts:
\begin{eqnarray}
\nonumber
\label{equatio6}
\textrm{I}(a,b)
& = &
\int_{-a}^{+a}dx
\Big[
G(x+b)  \, Q(-b)
-G(x+a) \, Q(-a)
+\int_{-b}^{-a} dy \, G(x-y) \, Q^{\prime}(y)
\Big] \, Q(x)
\\ 
& + &
\int_{-a}^{+a}dx
\Big[
G(x-a)  \, Q(a)
-G(x-b) \, Q(b)
+\int_{+a}^{+b} dy \, G(x-y) \, Q^{\prime}(y)
\Big] \, Q(x)
\end{eqnarray}
In the limit 
$\lim_{a \rightarrow + \infty}[\lim_{b \rightarrow + \infty} \textrm{I}(a,b)]$
the integrals and the terms depending on $b$, inside the square brackets of 
(\ref{equatio6}), vanishes because $\lim_{x \rightarrow \pm \infty} G(x)=0$.
For the terms depending on $a$ we can write:
\be
\lim_{a \rightarrow +\infty}
\int_{-a}^{+a} dx \, [\mp G(x\pm a) \, Q(\mp a) \, Q(x)]
=
\mp \frac{1}{2} \, Q^{2}(\mp \infty) \int_{-\infty}^{+\infty} dx \, G(x)
\ee
where we have used the fact that $G(x)$ is symmetric and peaked around the origin
together with the assumption that $Q(x)$ tends to some asymptotic value in
the limit $x \rightarrow \pm \infty$.
The generalization to the multicomponent case \cite{fsPZJ1} is straightforward.


\end{document}